\newcommand{\scell}[1]{$\mbox{#1}^*$-cell}
\newcommand{\scells}[1]{$\mbox{#1}^*$-cells}

\newcommand{\One}{{\bf 1}}
\newcommand{\ag}{\alpha}
\newcommand{\bg}{\beta}
\newcommand{\cg}{\gamma}
\newcommand{\dg}{\delta}
\newcommand{\sg}{\sigma}

\newcommand{\n}{\nu}
\newcommand{\eg}{\epsilon}
\newcommand{\lam}{\lambda}

\newcommand{\Sg}{\Sigma}
\newcommand{\Cg}{\Gamma}
\newcommand{\Dg}{\Delta}

\newcommand{\di}{\partial}
\newcommand{\be}{\begin{equation}}
\newcommand{\ee}{\end{equation}}
\newcommand{\bearr}{\begin{eqnarray}}
\newcommand{\eearr}{\end{eqnarray}}

\newcommand{\Real}{{\bf R}}

\newcommand{\bfg}{{\bf g}}

\newcommand{\threej}[6]{\protect{\scriptsize\mbox{$\left( \begin{array}{ccc}
      #1 & #2 & #3    \\      #4 & #5 & #6
    \end{array}     \right)$}}}

\documentstyle[12pt,psfig]{article}
\begin{document}

\title{A lattice worldsheet sum for 4-d Euclidean general relativity}
\author{Michael P. Reisenberger\\
 Instituto de F\'{i}sica, Universidad de la Rep\'{u}blica\\
      Trist\'{a}n Narvaja 1674, 11200 Montevideo, Uruguay}
\maketitle

\begin{abstract}
A lattice model for four dimensional Euclidean quantum general relativity is proposed for
a simplicial spacetime. It is 
shown how this model can be expressed in terms of a sum over worldsheets of spin networks in the 
lattice, and an interpretation of these worldsheets as spacetime geometries is given, based
on the geometry defined by spin networks in canonical loop quantized GR. The spacetime geometry has
a Planck scale discreteness which arises "naturally" from the discrete spectrum of spins of $SU(2)$ 
representations (and not from the use of a spacetime lattice). 

The lattice model of the dynamics is a formal quantization of the classical
lattice model of \cite{Rei97a}, which reproduces, in a continuum limit, Euclidean general 
relativity.
\end{abstract}

\section{Introduction}

The present work aims to provide a step in the construction of a theory of "quantum general
relativity" (QGR), meaning a theory within the framework 
of standard quantum mechanics\footnote{Except, perhaps, in the role played by 
time.} 
the classical limit of which is general relativity (GR), and in which the four
dimensional diffeomorphism invariance of GR is realized exactly, without 
quantum anomalies.

This conservative approach to the quantum gravity, in which one attempts to 
quantize GR without modification at the classical level or unification with 
other fields, has been revived by Ashtekar's discovery \cite{Ashtekar86}
\cite{Ashtekar87} of convenient new variables for classical canonical GR,
and the application of loop quantization \cite{Gambini86} to Ashtekar's 
canonical theory by Rovelli and Smolin \cite{RSloops}.


In Ashtekar's canonical theory the canonical variables are the left-handed 
(self-dual)\footnote{
In euclidean
GR the frame rotation group is $SO(4)$ which can be written as the product
$SU(2)_R\otimes SU(2)_L$. Left handed tensors 
transform only under the $SU(2)_L$ factor. Examples are left handed 
spinors and self-dual antisymmetric tensors, i.e. tensors $a$ that satisfy
$a^{[IJ]} = \eg^{IJ}{}_{KL} a^{KL}$.}
 part of the spin 
connection on 3-space and, conjugate to it, the densitized dreibein. The 
connection can thus be taken as the configuration variables, opening the door
to a loop quantization of GR. 

In loop quantization one supposes that 
the state can be represented by a power series\footnote{%
The series is not assumed to be 
convergent. A divergent series still defines a distribution on the space of generalized
connections via the Ashtekar-Lewandowski \cite{ALMMT95} inner product.}
in the spatial Wilson loops of 
the connection (which coordinatize the connections up to gauge), so the 
fundamental excitations are loops created by the Wilson loop operators.

At the present time the kinematics of loop quantized canonical GR 
is fairly well understood (see \cite{ALMMT95} for a recent review). That is to say, 
the space of states invariant under 3-diffeomorphisms of 3-space has been identified.
The kinematics alone leads to the striking prediction that geometrical observables
measuring lengths \cite{Thiemann_length}, areas \cite{discarea2}\cite{AL96a}
\cite{Fritelli96}, and volumes \cite{discarea2},\cite{AL96b} have discrete spectra 
and finite, Planck scale, lowest non-zero eigenvalues. 

However, the dynamics of the theory, is not well understood. 
The dynamics of QGR is encoded in the restrictions placed on 
physical states by the requirement of full {\em 4-diffeo} invariance. 
These restrictions are represented in the classical theory by the scalar, 
or "Hamiltonian", constraint (which, when formally quantized yields the 
Wheeler-deWitt equation).
\footnote
{In fact since the loop quantization of GR remains incomplete in the sense that no satisfactory 
quantum dynamics has been found, it remains possible that no such quantum dynamics even {\em can} 
be accomodated within the loop kinematics. If this is the case the whole loop quantization approach
to QGR would have to be abandoned, and with it the kinematical prediction of a discrete spectrum 
of areas.}

Thiemann \cite{QSD} has recently proposed a rigorously defined scalar constraint operator
in loop quantized GR. However, it seems that the theory defined by this constraint does
not have GR as its classical limit \cite{Smolin96}, \cite{GLMP97}. 

In a theory defined in terms of path integrals over 4-diffeo equivalence classes of histories
4-diffeo invariance is incorporated from the outset. Can quantum GR be
set up within the kinematical framework of loop quantization with transition amplitudes
defined by sums over 4-diffeo equivalence classes of histories? In addition to manifest 4-diffeo
invariance such a theory would incorporate the Planck scale discreteness of geometry implied
by the loop kinematics, which might provide a physical UV cutoff, both for the gravitational
field itself and any matter fields coupled to it. 

Here a lattice path integral model of loop quantized four dimensional Euclidean gravity
is proposed as a step toward such a theory. The model is a formal quantization of the classical
lattice model of \cite{Rei97a}, which reproduces, in a continuum limit, Plebanski's form of
Euclidean general relativity \cite{Plebanski77}.
The model (and a large class of others like it) can be formulated so that
the defining path integrals are sums over "spin worldsheets", the spacetime worldsheets
of spin networks. (Spin networks are graphs with edges and vertices carrying labels, and
are closely related to the loops of loop quantization. Each spin network embedded in space 
defines a state created by a certain finite polynomial of Wilson loops living on the graph,
and together these states span the state space of loop quantization.)

The advantages of a path integral formulation of loop quantization have been recognized
for some time \cite{Baez94},\cite{Rei94}. In particular, \cite{Rei94} proposes the representation
of quantum gravity as a sum over 4-diffeo equivalence classes of worldsheets of spin networks 
(spin worldsheets), and also the interpretation of these classes as discrete spacetime geometries.
What was needed was a way to translate the dynamics of gravity into this new framework.

Techniques for expressing lattice gauge theories in terms of sums over the worldsheets of
electric flux loops go back all the way to Wilson's strong coupling expansions \cite{Wilson74}.
In 1994 Iwasaki \cite{Iwasaki94,Iwasaki95} found a formulation of the Ponzano-Regge
model of 2+1 Euclidean GR in terms of a sum over the worldsheets of loops, and the author
found a formulation of arbitrary $SU(2)$ lattice gauge theories in terms of lattice spin 
worldsheets.\footnote{
Independnetly a worldsheet formulation of $U(1)$ gauge theories was found 
by Aroca, Baig, and Fort \cite{ABF94}, and the beginnings of such
a formulation for $SU(2)$ theories were developed by Aroca, Fort 
and Gambini \cite{AFG96}.}

Four dimensional Euclidean GR is an $SU(2)$ gauge theory when expressed in terms of 
self-dual variables, as in Ashtekar's canonical formulation, or Plebanski's covariant 
formulation \cite{Plebanski77}. Thus the methods of \cite{Rei94} could be used to translate 
a suitable lattice model of GR in terms of self dual variables into the spin worldsheet 
language. However, at the time no such lattice formulation of quantum GR existed.

In 1996 the author presented 
\footnote{
Seminars, Center for Gravitational Physics and Geometry, Penn State, April '96 and 
Erwin Shr\"odinger Institute, Vienna July '96.}
a simplicial model in terms of a sum over spins and $SU(2)$ link
variables to which the formalism of \cite{Rei94} can be applied directly. This "spin sum" model 
was obtained by a formal quantization of a classical simplicial action that reproduces Plebanski's 
formulation of Euclidean GR in a continuum limit \cite{Rei97a}.

These developments stimulated Rovelli and the author \cite{ReiRov} to construct a spin worldsheet 
sum for transition amplitudes from canonical loop quantized GR, using Thiemann's \cite{QSD} 
proposal for the Hamiltonian constraint. Such an approach has the advantage that one begins 
immediately in the continuum. However, the quantity they were able to express as a sum over
worldsheets, the exponential
of the Hamiltonian constraint with constant lapse, is more akin to evolution amplitudes of 
the gravitational field with respect to a physical clock 
than to the 4-diffeo invariant transition amplitudes considered here.

Markopolou and Smolin \cite{FotLee,Fot} have proposed a variant of the spin worldsheet formalism
which incorporates a Lorentzian causal structure intrinsic to the worldsheet and abandons the
topological spacetime as a home for the worldsheets.

Here the original spin sum model, and a hypercubic variant of it, are finally published. 
Recently, after the
work presented here was completed, some similar ideas have been presented in \cite{BaCr97} and 
\cite{Baez97}. In \cite{BaCr97} Barret and Crane propose a simplicial model in a similar vein 
to the one given here, with the interesting difference that GR is treated as an $SO(4)$ gauge 
theory. In \cite{Baez97}, by Baez, the proposal that GR be represented by a sum over 4-diffeo 
equivalence classes of spin worldsheets is presented carefully, and related to 2-category 
theory \cite{Crane_category}\cite{Baez_category}.

The present paper consists of two halves, the first dealing with the lattice and spin worldsheet
formalism in general, and the second with the particular models for quantum Euclidean GR being
proposed.

\S \ref{lattice} reviews the formalism of \cite{Rei94} from a somewhat modified perspective.
In \S \ref{local_lft} "local" $SU(2)$ lattice gauge theories, which can be represented 
by spin worldsheet sums, are defined. The procedure for calculating the worldsheet amplitudes for
such a lattice model is given in \S \ref{sws1}. \S \ref{cell:structure} and \S 
\ref{integration} deal with details omitted in the previous two subsections. The section closes
with \S \ref{geometint} which describes the geometrical interpretation of spin worldsheets.

In \S \ref{model} a lattice model for four dimensional Euclidean quantum general relativity 
of the type defined in \ref{local_lft} is proposed. \S \ref{model} presents the 
simplicial model, and \S \ref{motivation} motivates this model by showing how it is a formal 
quantization of the classical simplicial
model of \cite{Rei97a}. \S \ref{model_sws} discusses the spin worldsheet formulation of this
model. 

\section{Lattice gravity as a path integral over spin worldsheets}\label{lattice}

\subsection{Local lattice gauge theories of gravity} \label{local_lft}

In the class of lattice models we will consider spacetime 
is represented by a complex, $\Pi$, of four dimensional cells, each having
the topology of a 4-ball. $\Pi$ forms the "lattice", which need not be 
hypercubic or regular in any way, but which is a piecewise linear manifold. 

\begin{figure}
\centerline{\psfig{figure=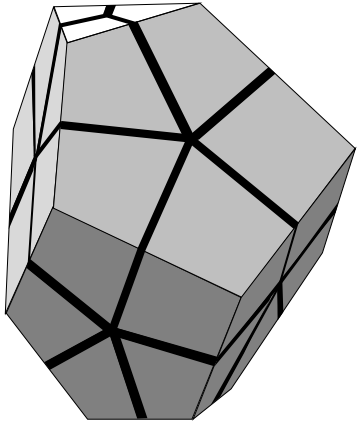,height=6cm}}
\caption[xxx]{The heavy black lines show the edges of the dual boundary 
$[\di\mu]^*$ of a 3-cell $\mu$. In a 4-cell, which is difficult to draw, the 
boundary is a three dimensional cellular complex, and the edges of the dual 
boundary connect the centers of the cells of this complex.}
\label{boundary_dual}
\end{figure}

The physical 
degrees of freedom in the different cells communicate via boundary data, 
which is required to match on the mutual boundaries of adjacent cells.
Specifically, the boundary data on a cell $\n$ is an $SU(2)$ lattice 
connection, consisting of $SU(2)$ parallel propagators along the edges of
a lattice $[\di\nu]^*$ on $\di \nu$, which is dual to $\di \nu$ seen as a 
three dimensional cellular complex. I will call $[\di\n]^*$ the 
``dual boundary'' of $\n$. ($[\di \nu]^*$ is illustrated in Fig. 
\ref{boundary_dual}). Because the cells communicate only via boundary data I
call these models ``local''. In the lattice gravity models we will consider 
the $SU(2)$ lattice connection serves as a discrete analog of Ashtekar's 
$SU(2)$ connection for Euclidean GR.\footnote{
See \cite{Fot} for an 
application of a related formalism to the the ``causal'' evolution scheme
of \cite{FotLee}.}

A quantum dynamical model within this framework is characterized by an 
$SU(2)$ gauge invariant quantum amplitude $a_\n(\bfg_{\di\n})$ for the
connection $\bfg_{\di\n}$ on the cell boundary. This 
amplitude can be thought of as the Hartle-Hawking state for a spacetime consisting 
of one cell only. In the path integral formalism it is the exponential of the action 
for the cell with the given boundary data.

The simplicial model of \cite{Rei97a} provides an action for 4-simplex cells 
which can be exponentiated to yield such a cellular amplitude. The action of 
a cell in \cite{Rei97a} in fact depends on several variables in addition to 
the connection on the boundary. However the connection is the only boundary 
data, i.e. the only data which is required to match between
neighboring cells, so the exponential of the cell action could be integrated 
over the remaining, internal, variables to yield an amplitude depending only 
on the boundary connection.

The continuum limit of the classical simplicial model of \cite{Rei97a} is 
Euclidean GR, in the sense that on sequences of simplicial histories that 
converge (in a sense defined in \cite{Rei97a}) to continuum gravitational 
field histories both the simplicial action and field equations converge to 
those of GR.\footnote{
In \cite{Rei97a} it is emphasized that the Regge model does {\em not} converge
to GR in this sense. I now think this may not reflect a real problem with
the Regge model. Solutions of the Regge model may simply converge to continuum
GR solutions in a weaker sense than that required in \cite{Rei97a}. Indeed,
W. Miller and Gentle \cite{MilGen97} observe this in the case of Kasner 
cosmologies.}
\footnote{
The model of \cite{Rei97a} approximates the Plebanski form of GR 
\cite{Plebanski77}\cite{CDJM}, which is not couched in terms of the metric and 
which extends GR to certain degenerate spacetimes not allowed in the standard 
metric formulation.} 
Using the cell amplitude $a_\n$ obtained from this action ensures that 
the continuum limit of the naive classical limit of the quantum model is GR. 
(Here the naive classical limit is the classical theory whose solutions are the stationary
points of the quantum amplitude for histories).
Of course we really want the continuum classical theory to emerge in a quite 
different way in quantum theory of gravity. We want it to emerge, at least in 
some of the states which represent large universes, as the behaviour 
of expectation values of observables which probe the ``classical
domain'' of gravity, i.e. phenomena at scales much larger than the
Plank scale.\footnote{
In this statement it has been assumed that quantum gravitational corrections 
to expectation values go to zero as the Plank scale goes to zero relative to 
all other scales in the problem, and so that the behaviour of the 
gravitational field is classical at all length scales well above the Plank 
scale. It has, however, been suggested that, contrary to these {\em prima facie} 
reasonable expectations, quantum effects should also be important at 
cosmological scales \cite{Mottola}. The identification of the classical domain
is not trivial.}   
Nevertheless, 
the correctness of the continuum limit of the naive classical limit provides 
motivation for developing quantum models based on the classical model of 
\cite{Rei97a}.

Work is in progress to obtain a tractable expression for the amplitude 
obtained by simply integrating out the internal (non-boundary) degrees of 
freedom in the exponentiated classical cell action. However, preliminary 
results suggest that this amplitude is quite complicated.

Once the model has been specified by the choice of a cell amplitude $a_\n$ 
the (unnormalized) amplitude $A(\bfg_{\di\Pi})$ for the connection 
$\bfg_{\di \Pi}$ on 
the boundary of the whole spacetime cellular complex $\Pi$ is obtained by 
multiplying together the amplitudes for all the individual 4-cells and then 
integrating over the connection on the mutual boundaries of the cells, i.e in 
the interior of $\Pi$, using the Haar measure to integrate over $SU(2)$ group 
elements:
\be                        \label{connection_p_int}
A(\bfg_{\di\Pi}) = \prod_{e \in X}\int d g_e 
                                   \prod_{\n\ \mbox{\scriptsize 4-cell of}\ \Pi} a_\n,
\ee
where $X$ is the set of connection bearing edges in the interior of $\Pi$, 
and $g_e$ is the parallel propagator along the edge $e$.

The connection, being the boundary data, has to match on the mutual
boundaries of cells. An important detail is that this makes it necessary to 
specify a parallel propagator ($SU(2)$ group element) for each of {\em half} 
of each edge of $[\di\nu]^*$, because half edges of $[\di\nu]^*$, not whole 
edges, in the boundaries of neighboring cells overlap. (Recall 
Fig. \ref{boundary_dual}.) More precisely: an edge $t$ of $[\di\nu]^*$ runs 
from a vertex inside one 3-cell of $\di\nu$ to another vertex inside a 
neighboring 3-cell. The half $t^{(-)}$ is the part of $t$ inside the 3-cell 
in which $t$ begins, and $t^{(+)}$ is the half in the 3-cell where $t$ ends.
In the complex $\Pi$ neighboring 4-cells share a 3-cell of their boundaries (we
will suppose each pair of neighbors shares only one 3-cell), so
the half edges of their dual boundaries in that 3-cell coincide. 
Matching the connection therefore requires
the parallel propagators along these half edges to match.\footnote{
Specifying the parallel propagators on the half edges of course gives no more 
$SU(2)$ gauge invariant information than specifying the parallel propagators 
on the whole edges. Indeed the gauge invariant content of the requirment that
the connection on mutual boundaries match can be expressed entirely in 
terms of the parallel propagators of entire dual boundary edges. One requiers
the triviality of the holonomies around certain curves formed from dual 
boundary edges - specifically, the star shaped curves formed by following, in 
turn, each dual boundary edge that is incident on a given 2-cell in $\Pi$ 
(see Fig. \ref{dual2cell} a ). This gauge invariant condition implies that the 
connections on the cell boundaries are gauge equivalent to connections that
match on mutual boundaries.} 

Clearly the amplitude $A$ that results from the integration is a function of 
the $SU(2)$ elements on all those dual cell boundary half edges that live in 
the boundary of $\Pi$ and thus are not integrated over. These half edges on 
$\di \Pi$ together form the dual to $\di \Pi$, so, just as for a single cell, 
the boundary data for the whole complex $\Pi$, consists of the parallel 
propagators along the (half) edges of the dual $[\di \Pi]^*$ to the boundary.

\subsection{Spin worldsheet formulations}   \label{sws1}

In this subsection the worldsheet formalism of \cite{Rei94} is reviewed from a
different perspective, that takes the cell amplitudes of the last section as
the starting point. (This approach takes elements from those of \cite{Iwasaki94, Iwasaki95}
and \cite{ReiRov}). The resulting formalism is entirely equivalent to that of \cite{Rei94}. 

The integral (\ref{connection_p_int}) over the connection on the interior of $\Pi$ 
that yields the amplitude $A({\bf g}_{\di\Pi})$ is just a discrete version of a path integral 
over connections. This path integral may be transformed into a sum over 
"spin worldsheets", which are the worldsheets of spin networks.

Spin networks (s-nets) are graphs with oriented edges carrying non-zero spins 
$j \in \{ \frac{1}{2},1, \frac{3}{2}, ...\}$ \footnote{
The edges carry spins
in $SU(2)$ spin networks. In spin networks of another group, $G$, the
edges carry irreducible representations of $G$.}
and vertices, with ordered incident edges, carrying ``intertwiners''. 

Intertwiners are $SU(2)$ invariant 
tensors.\footnote{
They are analogous to $\dg_{ij}$, $\eg_{ijk}$ and their products, which are 
invariant tensors of proper $SO(3)$.}
An intertwiner for a given vertex has, for each incident edge $e$, an index
of spin $j_e$. That is, the intertwiner is a vector of the tensor product
space formed from the representation spaces associated with each of the 
incoming edges. For trivalent vertices all intertwiners are proportional to
the Wigner $3-jm$ symbol $\threej{j_1}{j_2}{j_3}{m_1}{m_2}{m_3}$ (essentially 
a Clebsch-Gordan coefficient. See e.g. \cite{Yutsis}), but
for higher valence vertices the space of intertwiners is generally 
multidimensional. 

An s-net embedded in a space with connection defines a 
spin network {\em function}: a function of the connection which can be thought
of as a generalized Wilson loop. Then, given an embedded spin network $\Gamma$, 
the corresponding spin network function is obtained as follows.
\begin{enumerate}
   \item To each edge $e$ of $\Gamma$, carrying spin $j_e$, associate the
spin $j_e$ representation matrix of the parallel propagator along $e$. Note
that this matrix, $U^{(j_e)\,m}{}_{n}$ has one index living at each end of
the edge $e$. $m$ lives at the beginning and $n$ at the end.
 \item Contract the indices of the edge parallel propagators with
the corresponding indices of the intertwiners at the vertices at the ends
of the edges. If an edge is a closed loop the indices of the parallel 
propagator are contracted with each other, yielding a Wilson loop.
\end{enumerate}

s-net functions can be written as finite polynomials of Wilson loops
and span all such polynomials. For this reason they span the kinematical Hilbert
space of loop quantization \cite{RSnet,Baeznet,Fox95,Rei94,inv_loop,RdP}. 
On a finite lattice s-net functions span the whole 
space of gauge invariant distributions of the lattice connection.\footnote{
Any gauge invariant distribution $f$ on $C^\infty$ functions of the lattice
connection has an expansion in terms of s-net functions which converges
distributionally to $f$. That is to say, the series obtained
by integrating a test function $\phi$ against each term in the s-net 
expansion converges to $f[\phi]$. Distributions that are defined on a larger
class of functions of the connection form a subset of distributions on the
$C^\infty$ functions, so they also have distributionally convergent expansions
in terms of s-net functions. (On distributions see \cite{Courant}).}
A linearly independent basis s-net functions can be defined by for each unoriented graph 
a conventional orientation for each edge, and ordering of incident edges at each vertex,
and then choosing a linearly independent basis of intertwiners for each vertex of each graph.
The basis elements are then labeled by unoriented graphs, carrying spins on each edge, and
the name, or label, of a basis intertwiner at each vertex. If the intertwiner bases
are chosen orthonormal\footnote{
Here ``orthonormal'' means orthonormal with respect
to the inner product $(a,b) = \sum_{m = -j}^j a^*_m b_m$ ($m$ is incremented 
in integer steps in the sum).} 
and the s-net functions are multiplied by a normalizing factor $\sqrt{2j_e + 1}$ for
each edge $e$,\footnote{
A closed loop is treated as a chain of open edges joined at bivalent vertices,
which have normalized intertwiner $W^m{}_n = \frac{1}{\sqrt{2j + 1}}\dg^m{}_n$. The
spin network function of a closed loop is thus just the spin $j$ Wilson loop
with no further normalizing factor.}
with spin $j_e$, then the basis of s-net functions is orthonormal with 
respect to the natural Ashtekar-Lewandowski inner product \cite{ALinnerp}. On a lattice 
this inner product on functions of the connection reduces to
\begin{equation}
\langle \chi | \theta \rangle = 
\prod_{\mbox{\scriptsize $l$ edge of lattice}} \int dg_l\ \chi^*\: \theta,
\end{equation}
where the Haar measure is used to integrate over group elements. From here on only
orthonormal s-net bases will be considered.

The basis s-net functions are closely analogous to the momentum eigenstates 
of a particle, and just as momentum can be used instead of position as the 
boundary data for the motion of a particle, so a basis spin network on the boundary
of spacetime can be used in place of a connection as the boundary data for 
the gravitational field, at least in our class of lattice models.

The amplitudes of basis s-nets on the boundary $\di\Pi$ are the coefficients
of the corresponding s-net functions in the expansion of 
$A(\bfg_{\di\Pi})$. Using the orthonormality of the s-net basis these
amplitudes can be obtained as
\begin{eqnarray}
A(\Gamma) & = & \langle \chi_\Gamma | A \rangle \\
   & = & \prod_{\mbox{\scriptsize $t$ edge of $[\di\Pi]^*$} }\int dg_t\ 
\chi_\Gamma^*\: A.         \label{A_path_int}
\end{eqnarray}
Putting this
together with our earlier path integral prescription (\ref{connection_p_int}) 
for $A(\bfg_{\di\Pi})$ we
see that $A(\Gamma)$ can be calculated by multiplying $\chi_\Gamma^*$ with
the cell amplitudes $a_\n(\bfg_{\di\n})$ of all the cells of $\Pi$, and 
integrating over the connection on {\em all} of $\Pi$, including the boundary
$\di\Pi$.

If we now expand the amplitude, $a_{\nu}({\bf g}_{\di\nu})$, of the 
boundary connection of {\em each cell}, $\nu$, in s-net basis functions
we obtain an expression for the amplitude $A(\Gamma)$ which is both 
a sum over basis s-nets, and an integral over the connection. It turns out
to be quite easy to carry out the integration over connections in each term 
of the sum.\footnote{
No reversal of the order of integration and summation is needed,
because the spin network expansions of the cell amplitudes 
$a_\n(\bfg_{\di\n})$ converge {\em distributionally}. That is to say, the 
integral of a function $\phi$ against $a_\n$ is the sum of the integrals of 
$\phi$ against the terms in the expansion of $a_\n$. 
Thus, though we might hueristically think of the expansions of the cell 
amplitudes as pointwise convergent, so that we should sum them up to get the
integrand in (\protect{\ref{A_path_int}}) and then integrate over connections to
obtain $A(\Gamma)$, the reverse is in fact true: to get $A(\Gamma)$ we must
integrate over connections in each term in the expansion of the integrand of
(\protect{\ref{A_path_int}}) and then sum the resulting integrals. This, correct,
ordering of integration before summation is of course precisely the starting 
point for the spin worldsheet formulation I am about to describe.}
The result is a sum over spin worldsheets $S$ for $A(\Gamma)$:
\be              \label{sws:sum}
A(\Cg) = \sum_{S,\:\di S = \Cg}  w(S).
\ee

\begin{figure}
\centerline{\psfig{figure=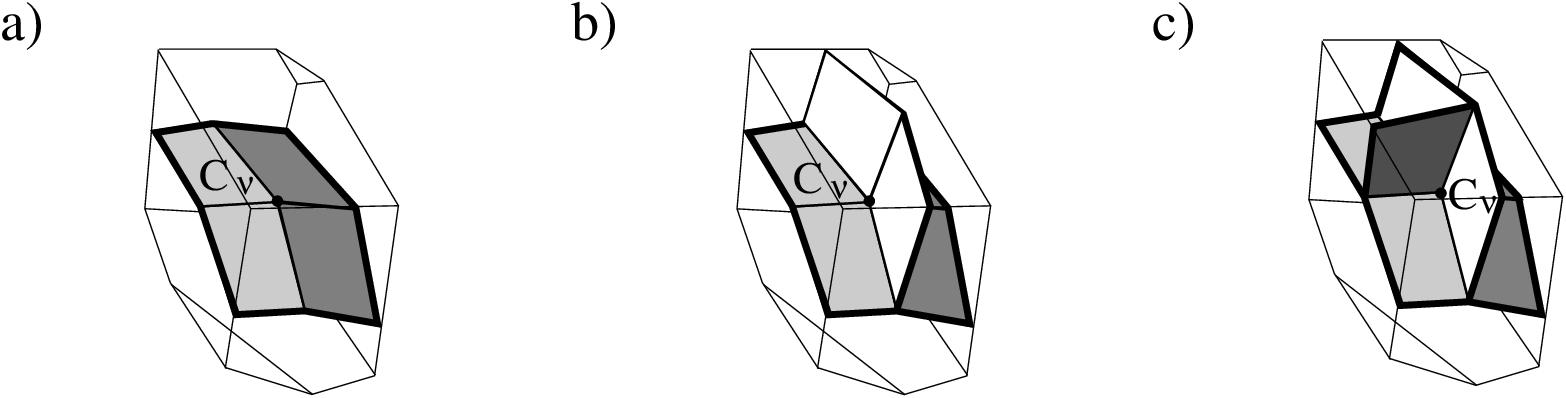,height=3.5cm}}
\caption[xxx]{The three panels show the spin worldsheets inside a cell 
spanning various types of spin networks on the cell boundary. 
(The spin networks are indicated by heavy lines). Since the four dimensional
situation is hard to draw three dimesional analogs are shown.\newline
\newline
Panel a) shows spin network consisting of a single unknotted loop
of spin $j$, which is spanned by a disk of spin $j$.\newline
\newline
Panel b) shows a spin network with two vertices and three edges.
It is spanned by three faces with the topology of disks which are joined
at a trivalent branch line running between the two vertices via
the center of the cell. Each face is bounded by one of the edges, and carries 
the spin of that edge. The two halves of the branch line, on either side of
the center, each carry the intertwiner label of the adjacent spin network
vertex. These can in general be distinct. 
\newline
\newline
Panel c) shows a spin network with four vertices and six edges. The four 
branch lines each start at a spin network vertex and end at the center of the 
cell, which serves as a branch point, or worldsheet vertex.}
\label{cellworldsheet1}
\end{figure}

\begin{figure}
\centerline{\psfig{figure=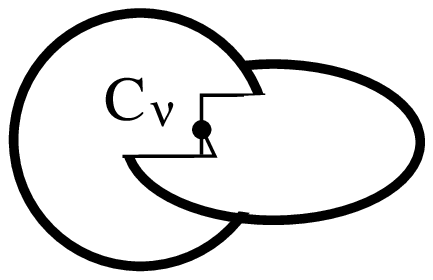,height=3.5cm}}
\caption[xxx]{A spin network consisting of two linked loops, and the spin
worldsheet that spans it are shown. Since loops cannot be linked in the 
topological 
2-sphere that is the boundary of a 3-cell, I cannot illustrate this type
of worldsheet with a three dimensional analog. Instead I have tried to
represent the four dimensional situation directly. In depicting knots in
a plane one uses breaks in the lines to indicate the parts of the lines
that are pushed down below the plane because there is a crossing. Here a 
broken surface in three dimensions is used to represent a surface in four 
dimensions, where the breaks indicate regions that are pushed into the fourth
dimension, off the 3-space which the viewer is visualizing (see \protect{
\cite{Carter}}
for more on this and other techniques of four dimensional visualization). 
We see that the worldsheet consists of two disks, each spanning a loop, which
intersect at a single point - the center of the cell. No attempt has been made
to show the 4-cell. The 3-space which the picture images is a 3-surface that 
cuts through the 4-cell in such a way that it contains most of the spin worldsheet.}
\label{cellworldsheet2}
\end{figure}

Let's first define spin worldsheets within the context of a single 4-cell $\n$.
A spin network on $\di\n$  which consists of a single 
loop carrying spin $j$ is spanned by a very simple spin worldsheet, namely
a disk in $\n$ carrying spin $j$. For an arbitrary spin network 
on $\di\n$ (consisting of edges of $[\di\n]^*$) a spin worldsheet can always
be constructed by taking the spin network and shrinking it continously to 
point $C_\n$ in the interior of $\n$. I will call the point $C_\n$, which may
be freely chosen, the ``center'' of $\n$.
The 2-surface swept out by the shrinking spin 
network forms the spin worldsheet, where each patch of worldsheet carries
the spin of the edge that swept it out, and each branch line carries the 
intertwiner
label of the vertex that swept it out. (The swept out surface can, and will, 
be chosen to have no self intersections). 
The spin worldsheets corresponding to different classes of spin networks on 
$\di\n$ are illustrated in Figs. \ref{cellworldsheet1} and 
\ref{cellworldsheet2}. 
In all cases spin 
worldsheets in a cell consist of one or more faces with the topology of disks, 
each carrying its own uniform spin, which are joined along branch lines 
carrying intertwiner labels.  

To each spin network on a cell boundary corresponds a particular cell spin 
worldsheet (modulo diffeomorphisms in the cell), and, clearly, each such
spin worldsheet corresponds to a unique spin network. Thus the sum over
cell boundary spin networks in the expression for $A(\Gamma)$ can be 
interpreted just as well as a sum over the possible assignments of cell spin
worldsheets to the cells of $\Pi$.

When the integral over connections is carried out it turns out that a 
non-zero contribution is obtained only from those terms in the sum in which all
the cell spin worldsheets together form a continous surface. Moreover, 
spins and (if suitable intertwiner bases are used) intertwiner labels on
this surface must match across boundaries between cells. At $\di\Pi$
the spins and intertwiner labels on the surface must match those of the 
boundary spin network $\Gamma$. 

\begin{figure}
\centerline{\psfig{figure=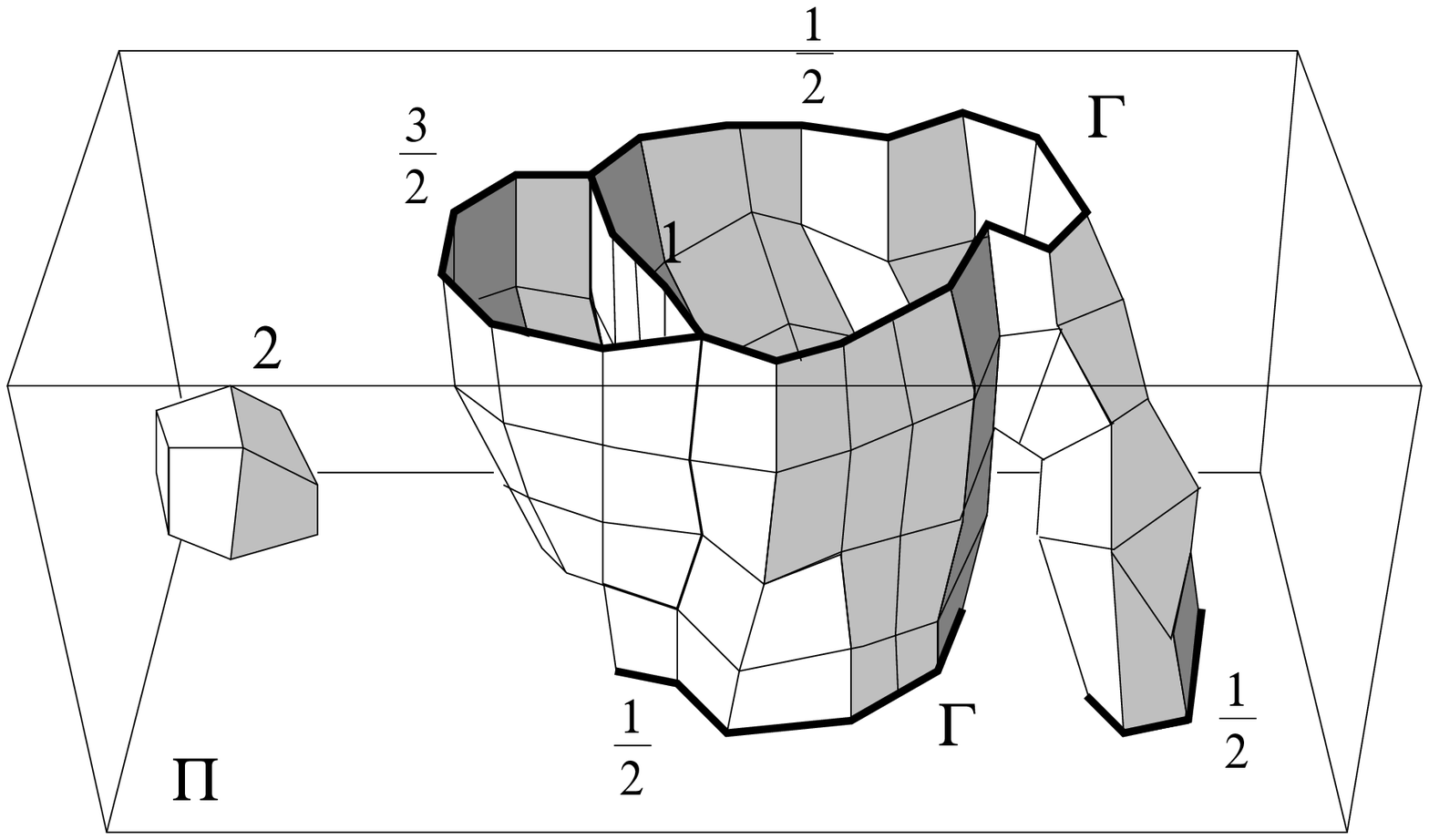,height=5cm}}
\caption[xxx]{The diagram shows a lattice spin worldsheet in a three dimensional 
spacetime. The worldsheet consists of four unbranched components, three
of which meet along a branch line, while the fourth forms an isolated
bubble. The three connected unbranched components also meet the boundary of
the spacetime $\Pi$ on a spin network $\Gamma$, which is drawn with heavy 
lines. A possible assignment of spins to the unbranched components and the 
edges of $\Gamma$ have been written in. Note that the branch line carries no 
intertwiner label since it is only trivalent. To keep the figure simple
only the boundary $\di\Pi$ of the spacetime cellular complex has been shown,
and that has been chosen to be a simple box, even though $\di\Pi$ can, in 
fact, be quite irregular. The relationship of the spin worldsheet to the 
individual cells of $\Pi$ is illustrated in Figs. \ref{cellworldsheet1} and
\ref{cellworldsheet2}. 
The spin worldsheet shown is made up of quadrangles. This is in fact generally
true, as will be explained further on in the text. However, the quadrangles
will generally not form a rectangular grid. Some possible features of 
worldsheets that are not illustrated are self 
intersections at points and non-orientable components. }
\label{grspinws}
\end{figure}

I shall call the union of the cell spin worldsheets simply ``the spin 
worldsheet''. This surface will generally have branch lines and self 
intersection points, preventing it from being a true 2-manifold. 
(See Fig. \ref{grspinws}). However, it can always be decomposed into
unbranched components, bounded by edges of $\Gamma$ or branch lines, at
which three or more unbranched components meet. These unbranched components
may have transverse intersections at isolated points, including self intersections, 
but otherwise they are 2-manifolds.
The matching conditions across inter cell boundaries and at $\di\Pi$ show
that each unbranched component carries uniform spin, and that those 
unbranched components that meet the boundary $\di\Pi$ are bounded there by
edges of $\Gamma$ carrying the same spin. The matching conditions also ensure 
that on $\di\Pi$ the intertwiner labels on branch lines match those of the 
vertices of $\Gamma$. They leave open the possibility that the intertwiner 
labels may change at the centers of cells.\footnote{
In lattice Yang-Mills and BF theories, which can be formulated within
the present framework \cite{Rei94}, one may choose orthonormal intertwiner 
bases for the various types of spin network vertices once and for all, and 
the intertwiner labels will be constant along branch lines.} 

Since the integral over connections in the expression for $A(\Gamma)$ is 
non-zero only for those assignments of basis s-nets
to the cell boundaries which correspond to spin worldsheets, 
$A(\Gamma)$ can be represented as a sum over spin worldsheets, with the weight
of each worldsheet given by the value of the integral over connections
for the corresponding assignment of cell boundary spin networks.
   
The name spin worldsheet is justified for the surface I have defined
because, firstly, it spans the boundary spin network, and secondly,
any cross section of the spin worldsheet by a lattice hypersurface, i.e.
a 3-surface made of cell boundaries, is a spin network, and thus a legitimate
intermediate state in the history of an evolving spin network. The cross 
sections are spin networks because the edges of a cross section, being
each a cross section of an unbranched component, carries constant spin, and
because the vertices are vertices of cell boundary spin networks, and thus
allowed spin network vertices.\footnote{
For $SU(2)$ spin networks the spins of edges incident at a vertex must satisfy
the polygon condition: the spins must be the edge lengths of a polygon of
integer circumference which can be realized in the plane \cite{Yutsis}.}

That completes the introduction to local lattice gauge theories and 
their formulation in terms 
of a path integral over spin worldsheets. To be able to work with these ideas 
requires a more thorough understanding. In the following two subsections I
will return to the main elements of the preceeding discussion and develop them
more fully. Specifically, I examine the spacetime cellular structure used
(\S \ref{cell:structure}), and the integration over connections that yields 
the worldsheet sum for $A(\Gamma)$ (\S \ref{integration}).

\subsection{Cellular structure and lattice connections in more detail} 
\label{cell:structure}

Let me begin by describing the cellular complex $\Pi$ more precisely.
$\Pi$ consists of 4-dimensional cells $\nu$ each having 
the topology of a 4-ball. The boundaries of the 4-cells are made of 3-cells. 
These either belong to the boundary $\di \Pi$ of the complex or are shared by 
precisely two 4-cells. A pair of 4-cells may share at most one 3-cell, and 
similarly 3-cells may share at most one 2-cell, etc. 

\begin{figure}
\centerline{\psfig{figure=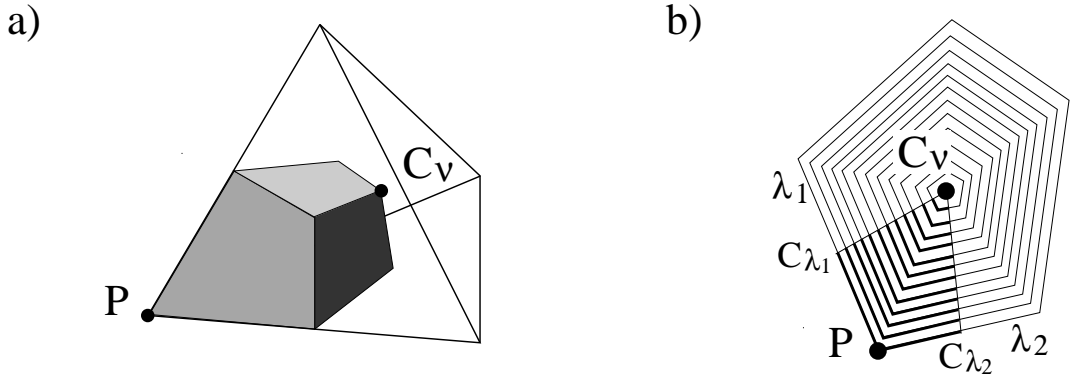,height=5cm}}
\caption[xxx]{Panel a) shows the corner cell $c_P$ associated with the vertex $P$ 
of a 3-simplex $\mu$. Notice that the intersection of $c_P$ with any of the 
triangular faces of $\mu$ that are incident on $P$ is itself the two 
dimensional corner cell of $P$ in the face in question. Note also that $c_P$ 
is diffeomorphic to a cube, and each of the subsimplices of $\mu$ that 
touch $P$ (including $\mu$ and $P$) contain one corner of $c_P$. These 
features are shared by corner cells in any cell (not necessarily a simplex).
\newline\newline
Panel b) shows a two dimensional example of the construction of corner cells 
within a generic polygonal cell $\mu$. The heavy line along the boundary 
$\di\mu$ shows the dual boundary cell $P^*_{\di\mu}$, formed by the union of 
the two corner cells of $P$ in $\di\mu$. The images
of $P^*_{\di\mu}$ in the concentric, sucessively smaller images of $\di\mu$ 
are also indicated by heavy lines. Together these sweep out the corner cell 
of $P$ in $\mu$. The 1-cells $\lam_1$ and $\lam_2$ of $\di\mu$ incident on $P$, 
mentioned in the text, and their centers are labeled.}
\label{cellular_substructure}
\end{figure}

In the interior of each $d$-cell $\mu$ we choose a point that we call its 
``center'', $C_\mu$.
Using the centers we will define a cellular substructure of $\mu$ 
(as illustrated in Fig. \ref{cellular_substructure}). $\mu$ will be divided 
into ``corner cells'', each containing one vertex (or ``corner'') of $\mu$. 
If $\mu$ is a 1-cell with endpoints $P$ and $Q$ the corner cell $c_{P\mu}$ is 
the segment $PC_\mu$. If $\mu$ is a 2-cell with a vertex $P$ shared by two 
1-cells $\lambda_1$ and $\lambda_2$ in $\di\mu$ then $c_{P\mu}$ is the 
quadrangle $C_\mu C_{\lambda_1} P C_{\lambda_2}$ (See Fig. 
\ref{cellular_substructure} b.)
For $\mu$ a cell of arbitrary dimensionality the decomposition into corner 
cells can be described as follows: each vertex $P$ of $\mu$ has a number of 
$d-1$ dimensional corner cells in $\di\mu$. If we let $P^*_{\di\mu}$ be the 
union of the corner cells of $P$ in $\di\mu$, then the corner cell of $P$ in 
$\mu$ is essentially the cone over $P^*_{\di\mu} \subset \di\mu$ with vertex 
at $C_\mu$. More precisely, since $\di\mu$ is 
contractible in $\mu$, which has the topology of a ball, one can define a 
continous family of diffeomorphic 
images of $\di\mu$ that converge on $C_\mu$ and cover $\mu$ exactly once. 
Each of these diffeomorphic images contains as a subset the image of 
$P^*_{\di\mu}$. The union of these images is the corner cell $c_{P\mu}$. 
This construction of corner cells is illustrated in Fig. 
\ref{cellular_substructure} b). Of course the construction only defines the 
cellular decomposition up to isotopy, but that is all we need.

\begin{figure}
\centerline{\psfig{figure=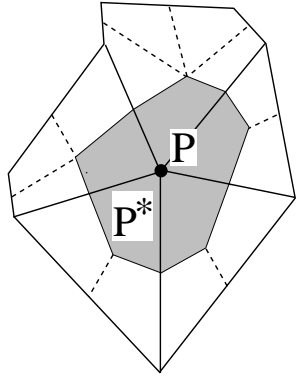,height=5cm}}
\caption[xxx]{The dual cell $P^*$ dual to the vertex $P$ in a two dimensional 
cellular complex is shown. The boundaries of the other dual cells are 
indicated by dashed lines. Note that the cells of the dual 
complex constructed this way generally are not flat (in a geometry that
makes the 4-cells of the original complex flat).}
\label{dualfig}
\end{figure}

Notice that the union $P^*_\Delta$ of corner cells of the vertex $P$ in the 
complex $\Dg$ forms a cell of a complex $\Dg^*$ dual to $\Dg$. See 
Fig. \ref{dualfig}. The dual $[\di \nu]^*$ of the boundary $\di \nu$ of a 
4-cell $\n$ consists of the 3-cells $P_{\di\nu}^*$ and the lower dimensional
cells derived from these. 

So far we have defined three four dimensional cellular complexes, our
original cellular complex $\Pi$, a complex $\Pi^*$ which is topologically 
dual to $\Pi$, and a finer complex called the ``derived complex'', $\Pi^+$, 
\cite{Maunder} built of four dimensional corner cells, which can be thought 
of as the intersections of 4-cells of $\Pi$ and $\Pi^*$. 
From here on a ``$d$-cells'', ``$d^*$-cells'', and ``$d^+$-cells'' will, unless 
otherwise qualified, be $d$ dimensional cells of $\Pi$, $\Pi^*$, and $\Pi^+$ respectively.

\begin{figure}
\centerline{\psfig{figure=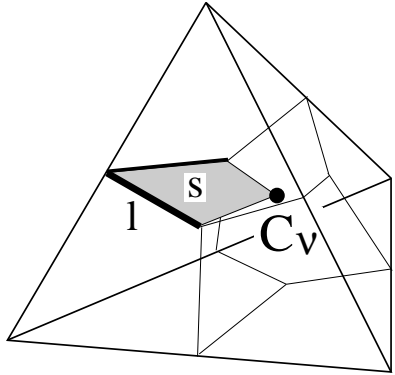,height=5cm}}
\caption[xxx]{To illustrate the idea of a wedge a wedge of a 3-cell (here a 3-simplex) is shown, though
the wedges that are really of interest are wedges of 4-cells. A dual boundary edge $l$ is indicated
by a heavy line. Note that this edge starts on a front face of the tetrahedron, but continues
on one of the back faces. The other dual boundary edges are indicated by thin lines. The shaded
plane region $s$ inside the tetrahedron is the wedge defined by $l$ and the center $C_\n$ of
the tetrahedron.} 
\label{wedges}
\end{figure}

Certain 2-cells of the derived complex, which I will call ``wedges'', play
a central role both in the building of models and in their spin worldsheet
formulation. Fig. \ref{wedges} shows a wedge in a 3-simplex.

Wedges are those 2-cells of the derived complex which touch the center of some
4-cell of $\Pi$. Each wedge is a quadrangle, with one corner at the 
center of its $\Pi$ 4-cell $\n$, and two sides on the boundary of $\n$. 
The intersection of the wedge with $\di\n$ is in fact an edge of the dual
boundary $[\di\n]^*$. (See Fig. \ref{wedges}) This follows from the fact that 
the corner cells are cones, with vertex $C_\n$, over the 3-cells of the dual boundary. 

Wedges are the basic building blocks of spin worldsheets. The cell spin 
worldsheets discussed earlier (see Figs. \ref{cellworldsheet1} and 
\ref{cellworldsheet2}) consist of the wedges 
in the given cell that are bounded by edges of the basis spin network on $\di\n$,
i.e. by those dual boundary edges that carry non-zero spin.
Each wedge carries the spin of its edge. 

\begin{figure}
\centerline{\psfig{figure=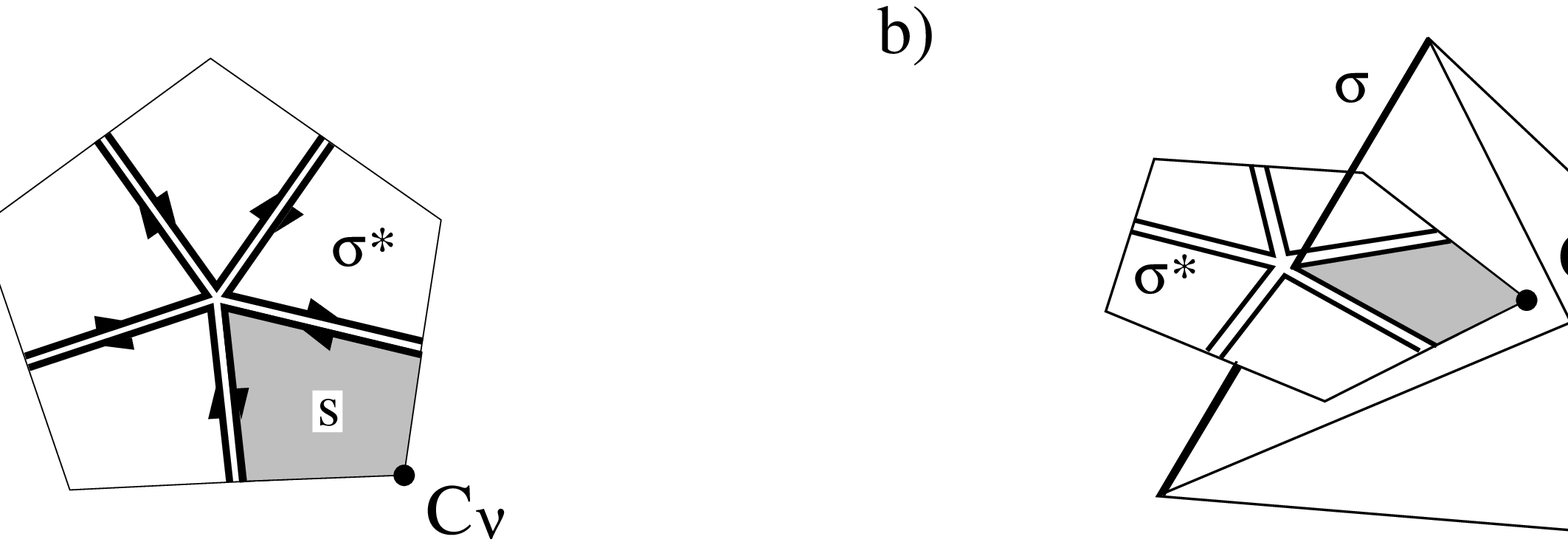,height=5cm}}
\caption[xxx]{Panel a) shows a \scell{2} $\sg^*$. $\sg^*$ dual to the 2-cell $\sg$ of 
$\Pi$, which passes through its center. The corners of $\sg^*$ are the centers 
of the 4-cells of $\Pi$ incident on $\sg$. The heavy lines show those
dual boundary edges belonging to the boundaries of these 4-cells that live 
on $\sg^*$. $\sg^*$ is the union of the wedges associated with these
dual boundary edges. In order to write spin network functions on the cell 
boundaries in terms of parallel propagator matrices we choose 
orientations for the dual boundary edges. A convenient choice, which will be 
used, is to choose an orientation for each \scell{2}, use this to define 
the orientation of the constituent wedges, and take the consequent orientation 
of the boundary of the wedge as the orientation of the dual boundary edge that
forms part of it. Such a choice of orientation is indicated by arrows on the 
dual boundary edges in the figure.
\newline\newline
Panel b) shows, in the context of a {\em three} dimensional complex, a 
\scell{2} $\sg^*$, the 1-cell $\sg$ that it is dual to, and one of the incident 
3-cells (a simplex $\n$). }
\label{dual2cell}
\end{figure}

Finally, it is helpful to note the relationship of the wedges with the dual complex 
$\Pi^*$: The union of all the wedges incident
on a 2-cell $\sg$ of $\Pi$ forms the 2-cell $\sg^*$ of $\Pi^*$ which is dual to 
$\sg$. This is illustrated in Fig. \ref{dual2cell}.\footnote{
The proof is not difficult.
One notes that a non-empty intersection $\n \cap P^*$ of a 4-cell $\n$ and a dual
4-cell $P^*$ is a corner cell. It follows that the portion $\sg^*\cap\n$ of a 
\scell{2} $\sg^*$ in $\n$ is a 2-cell in the boundary of a corner cell. 
Moreover, $\sg^*$ must touch $C_\n$, the only site of $\Pi^*$ in $\n$, and
$\sg$, the 2-cell it is dual to, so $\sg^*\cap\n$ must be the wedge that meets
$\sg$.} 

It follows from the fact that spin worldsheets can have no open boundaries in 
the interior of $\Pi$ that, if a wedge belongs to the
worldsheet, then the whole \scell{2} it belongs to must be part of the 
worldsheet. That is, spin worldsheets are made of entire dual 
2-cells. (Note that $\Pi^*$ has been defined so that the spacetime it covers, i.e. 
the union of its cells, is the same as that of $\Pi$: \scells{2} are therefore 
cut off beyond $\di\Pi$.)

A role will be also be played by the edges of $\Pi^+$ in the interiors of 4-cells.
In a given 4-cell, $\n$, these edges form the mutual boundaries of the wedges.
They connect the center, $C_\n$, of the 4-cell with the centers of its bounding 3-cells. 
The centers of the 3-cells are of course the vertices of the dual boundary $[\di\n]^*$.
Notice also that these edges are halves of \scells{1}.
If $\n$ and $\n'$ are two adjacent 4-cells, and $\tau$ is the 3-cell 
forming their mutual boundary, then the edges $C_\n C_\tau$ and
$C_{\n'} C_\tau$ together form the \scell{1} connecting $C_\n$ and $C_{\n'}$.

\subsection{The integration over connections}\label{integration}
 
Armed with the geometrical imagery of \S \ref{cell:structure} we are now 
ready to tackle the integration over the connection required to find the 
worldsheet amplitudes in the sum (\ref{sws:sum}) for $A(\Gamma)$. That is, 
we suppose that the cell amplitudes $a_\n$ in the integrand of (\ref{connection_p_int}) have 
all been expanded on spin network bases, and we now integrate one term in 
this expansion over connections. 

To make the math as clean as possible I will adopt particular conventions 
regarding the intertwiner bases and the orientations of the wedges. Firstly, 
I will require that all the wedges in a \scell{2} are coherently oriented. This 
induces opposed orientations in overlapping pairs of dual boundary
edges (see Fig. \ref{dual2cell} a). As a result, when we integrate over the parallel 
propagators along half dual boundary edges the only type of integral that 
ever appears is 
\be  \label{basic:integral}
\int_{SU(2)} U^{(j_1)}(g)^{m_1}{}_{n_1} U^{(j_2)}(g^{-1})^{m_2}{}_{n_2}\:dg = \frac{1}{2j_1 + 1}
\dg_{j_1 j_2} \dg^{m_1}_{n_2} \dg^{m_2}_{n_1}.
\ee
If $\n_1$ and $\n_2$ are two adjacent 4-cells then the edges and the vertices
of the dual boundaries $[\di\n_1]^*$ and $[\di\n_2]^*$ overlap in the 3-cell, 
$\tau$, shared by $\n_1$ and $\n_2$. The integral (\ref{basic:integral}) arises 
as the integral over the parallel propagator, $g$, along a half edge $r_1$ of
$[\di\n_1]^*$, lying in this 3-cell. Since the overlapping half edge $r_2$ of
$[\di\n_2]^*$ has the opposite orientation it has parallel propagator $g^{-1}$.

Note that the parallel propagator along an edge, in 
any representation, is a two point tensor that transforms as a vector under 
gauge transformations at the beginning of the edge, and as a co-vector under 
gauge transformations at the end (because it transports a vector from the 
end of the edge back to the beginning). The upper index $m_1$ in 
$U^{(j_1)\,m_1}{}_{n_1}$ therefore lives at the beginning of $r_1$, 
while the lower index, $n_1$, lives at the end. The indices of 
$U^{(j_2)}(g^{-1})^{m_2}{}_{n_2}$ are similarly housed with respect to 
$r_2$. Since $r_2$ is antiparallel to $r_1$ this  means that $m_2$ lives at 
the end of $r_1$, while $n_2$ lives at the beginning of $r_1$. The Kronecker deltas 
in (\ref{basic:integral}) thus each connect indices living at the same end of $r_1$. 


The second convention regards the intertwiner bases. I will 
choose the intertwiner bases at the overlapping vertices of $[\di\n_1]^*$ and
$[\di\n_2]^*$ in $\tau$ to be complex conjugates of each other. That is to say, if 
$\{ W_{1\,I}^{\bf j}\}$ is a basis of intertwiners for the vertex of
$[\di\n_1]^*$ (with ${\bf j} = [j_1, j_2, ...]$ the spins of the incident edges 
at the vertex, and $I$ a label identifying the distinct basis intertwiners 
for the same $\bf j$) then the basis intertwiners of the corresponding 
vertex of $[\di\n_2]^*$ are\footnote{
Upstairs indices are vector indices under gauge transformations, and in 
intertwiners correspond to incoming edges at the vertex, while downstairs 
indices are covector indices corresponding to outgoing edges of the vertex. 
Gauge transformations act on co-vectors with the inverse of the vector 
transformation matrix, so that the contraction of a vector and a co-vector 
is a scalar. Note that in the unitary representations of the gauge 
transformations we are using complex conjugation turns vectors into 
covectors, and {\em vice versa}, because, being unitary, the gauge 
transformations preserve the inner product \protect{$(a, b) = a^*\cdot b$}. 
It follows 
that complex conjugation turns upstairs indices into downstairs indices 
and {\em vice versa}.}
\be
W_{2\,I}^{\bf j}{}_{m_1 ... m_a}{}^{n_1 ... n_b} 
= [W_{1\,I}^{\bf j}{}^{m_1 ... m_a}{}_{n_1 ... n_b}]^*.
\ee
Note that this prescription is compatible with my orientation convention. 
Complex conjugation turns upstairs indices into downstairs indices, and 
{\em vice versa}, so if $W_1$ is an intertwiner for a vertex $v_1$, then 
$W_1^*$ is an intertwiner for a vertex like $v_1$, but with the orientation
of all incident edges reversed.
 
Since we use orthonormal intertwiner bases, in the sense that
\be
W_{1\,I}^{\bf j} \cdot [W_{1\,I'}^{\bf j}]^* = \dg_{I\,I'}
\ee
(where $\cdot$ signifies contraction on all indices) my convention implies that
\be  \label{inter:orthonorm}
W_{1\,I}^{\bf j} \cdot W_{2\,I'}^{\bf j} = \dg_{I\,I'}.
\ee
That is, the complete contraction of the intertwiners on two overlapping vertices 
of cell boundary s-nets with the same incident spins is $1$ if they carry the same 
basis intertwiner label $I$, and zero otherwise.

The integral over connections can now be carried out using 
(\ref{basic:integral}) and (\ref{inter:orthonorm}). All that needs to be 
done is to organize the integrations. I will organize them by 
\scells{2}. That is, I will do the integrations on all the dual
boundary edges living on the same \scell{2} together. There are two types of 
\scells{2} to consider: \scells{2} in the interior of $\Pi$, and \scells{2} 
bounded by $\di\Pi$.


Let's consider a $p$ sided \scell{2}, $\sg^*$, in the interior of $\Pi$. 
The integral is zero unless all the wedges of $\sg^*$ carry the same spin. 
If the wedges of $\sg^*$ {\em do} carry a common spin $j$, then the integration yields
\begin{itemize}
\begin{enumerate}
\item a Kronecker delta at each \scell{1} in $\di\sg^*$, i.e. on each \scell{1} a Kronecker 
      delta on the two dual boundary parallel propagator indices that live there;
\item a chain of contracted Kronecker deltas at the center of $\sg^*$, contributing, when evaluated,
       a factor $\sum_{m=-j}^{j} \dg^m_m = 2j + 1$;
\item a factor $(\frac{1}{2j + 1})^p$ due to the $\frac{1}{2j + 1}$ appearing in the integral
(\ref{basic:integral}) and the normalizing factors $\sqrt{2j + 1}$ associated with each
dual boundary edge.\footnote{
Each edge of a basis s-net function carries a normalizing factor
$\sqrt{2j+1}$. Because the normalized intertwiner for bivalent vertices
is $W^m{}_n = \frac{1}{\sqrt{2j + 1}} \dg^m{}_n$, the normalization resulting from this 
convention is unchanged if an edge is 
split into a chain of edges joined by bivalent vertices.}
$(\frac{1}{2j + 1})^p$ can be thought of as a factor of $\frac{1}{\sqrt{2j+1}}$ for each
\scell{1} in $\di\sg^*$.
\end{enumerate}
\end{itemize}

If $\sg^*$ is bounded by $\di\Pi$ the situation is essentially the same. The only difference is
that in this case the integral is non-zero only if the wedges {\em and} the edge of the dual 
boundary $[\di\Pi^*]$ that bounds $\sg^*$ all carry the same spin $j$. Thus, if $j\neq 0$
$\sg^*$ must be bounded by a spin $j$ edge of $\Gamma$.

Now let's focus on the vertices of the dual boundaries $[\di\n]^*$. These come in overlapping, 
or "facing", pairs. In the interior of $\Pi$ a 3-cell contains two overlapping vertices, each 
belonging to one of the two 4-cells sharing the 3-cell. 
It is convenient to think of the dual boundary $[\di \Pi]^*$, on which $\Cg$ lives, as a distinct
complex that overlaps the dual boundaries of the 4-cells that meet $\di\Pi$. Then 
a 3-cell in $\di\Pi$ contains a vertex belonging to the dual boundary of a 4-cell and a facing 
vertex belonging to $[\di\Pi]^*$.

In either case integration over the connection has replaced the parallel 
propagation matrices that are contracted with the vertex intertwiners 
in the spin network functions, with Kronecker deltas that contract the intertwiners
at facing vertices. (These are the Kronecker deltas at the edges of the 
\scells{2} in 1. in the list of factors given above.)

As already explained, the wedges in a \scell{2} must carry the same spin for the integral over 
connections to be non-zero. If this is the case the spins of the pair of incident overlapping edges 
at facing vertices are equal. (\ref{inter:orthonorm}) therefore implies that the contraction of the
facing intertwiners is $\dg_{I\,I'}$, where $I$ and $I'$ are the basis intertwiner labels at the
two vertices.

Two consequences can be drawn immediately from the above results. Firstly,
the fact that only spin assignments in which \scells{2} carry uniform spin 
contribute to $A(\Gamma)$ implies that the unbranched components of spin 
worldsheets consist of entire \scells{2}. Secondly, the basis intertwiner 
label on a branch line does not change
where the branch line crosses a 4-cell boundary.

To assemble all the factors and obtain an expression for $w(S)$ we now shift 
our focus to the 4-cells. Three basic types of spin worldsheets can occur 
inside a 4-cell (recall Fig. \ref{cellworldsheet1}): a disk; a collection of 
half disks joined, like the pages of a book, on a branch line crossing the cell; 
and branch line "vertices", in which three or more branch lines meet at the center 
of a cell. Since the cell is four dimensional it may also contain several worldsheets 
of these types, intersecting at the center of the cell. Two disks intersecting 
at a point are illustrated in Fig. \ref{cellworldsheet2}. I will refer to both intersections
and branch line vertices as vertices of the spin worldsheet.

Let me first show that each unbranched component of a spin worldsheet $S$ 
carries uniform spin, by considering disk type cell spin worldsheets.
Let $\n$ be a 4-cell containing a cell spin worldsheet 
$S\cap\n$ consisting of a disk, and possibly other components that intersect the disk at the
center of the cell. Recall that the centers of 4-cells are sites of the dual lattice $\Pi^*$. 
The part of the spin worldsheet inside a 4-cell $\n$ will have the topology of a disk iff $C_\n$ 
is a point on the interior of an unbranched component.\footnote{
Recall also that in our definition of unbranched components isolated intersection 
points don't count as branching. So such an intersection 
point can be an interior point of an unbranched component}
The spin network on $\di\n$ which bounds the disk is just a single loop, and thus carries a single
uniform spin. This implies that the disk, and thus all the \scells{2} 
incident on $C_\n$, must carry the 
same spin. Applying this argument to every interior point of an unbranched component then shows that
the whole component must carry uniform spin. 

Assembling all the factors that contribute to the amplitude of an unbranched component 
$\varsigma$ of spin $j$, we find: 
\begin{itemize}
\begin{enumerate}
\item from the integration over connections, a factor of $\frac{1}{\sqrt{2j + 1}}$ for each edge 
           of the boundary of each \scell{2} in $\varsigma$. Thus there is a factor of 
           $\frac{1}{2j + 1}$ for each interior \scell{1} of $\varsigma$, and a factor 
              $\frac{1}{\sqrt{2j + 1}}$ for each \scell{1} in $\di\varsigma$;
\item also from the integration over connections, a factor of $2j + 1$ for each \scell{2} of 
          $\varsigma$;
\item for each \scell{0} (center of a 4-cell) in the interior of $\varsigma$ the amplitude of 
           the cell boundary spin network $a_\n[\gamma]$, with $\gamma = \varsigma\cap\di\n$. 
\end{enumerate}
\end{itemize}
The amplitude $w(S)$ of the whole spin worldsheet $S$ is then the product of the amplitudes of
the unbranched components, times the cell amplitudes $a_\n$ of empty 4-cells, and of ones 
containing branch lines and vertices.

This result can be put in a more transparent form by working with "reduced amplitudes" for
spin networks and spin worldsheets, obtained from $A(\Cg)$, $a_\n(\cg)$,
and $w(S)$ by a change of normalization.
The reduced amplitude $\bar{A}(\Cg)$ of the spin network $\Cg$ on $\di\Pi$ is $A(\Cg)$ multiplied
by a factor $\frac{1}{2j + 1}$ for each closed loop in $\Cg$, and a factor 
$\frac{1}{\sqrt{2j + 1}}$ for each edge connecting vertices of valence $\geq 3$. The reduced
cell amplitude $\bar{a}_\n(\cg)$ is defined analogously, and $\bar{w}(S)$ is 
obtained by multiplying $w(S)$ by the normalizing factor associated with its bounding
spin network $S \cap \di\Pi$. Therefore,
\be
\bar{A}(\Cg) = \sum_{S,\di S = \Cg} \bar{w}(S).
\ee

When the product of the cell amplitudes in $\bar{w}(S)$ is rewritten in terms of reduced cell
amplitudes the factors of $2j+1$ stemming from the change to reduced amplitudes and those
stemming from the integration over the connection combine into, simply, a factor
\be
(2j_\varsigma + 1)^\chi[\varsigma]
\ee
for each unbranched component, where $\chi[\varsigma]$ is the Euler characteristic of $\varsigma$.
For any 2-surface of $\Pi^*$
\be
\chi = N_2 - N_1 + N_0,
\ee
where $N_i$ is the number of \scells{$i$} in the surface. As is well known the 
Euler characteristic of a surface depends only on its topology.

The reduced spin worldsheet amplitude can thus be written as
\be             \label{sws_amp1}
\bar{w}(S) = \prod_{\varsigma \mbox{\scriptsize unbranched component}}
          (2j_\varsigma + 1)^{\chi[\varsigma]}\ \prod_{\n\ \mbox{\scriptsize 4-cell of}\ \Pi}
             \bar{a}_\n(\cg[S]).
\ee

That is almost the end of the story.\footnote{%
The sign factors found here were missed in \cite{Rei94}.}
One detail remains: The reduced cell amplitudes
$\bar{a}_\n(\cg)$ depend on the orientations of the edges of $\cg$ and the ordering
of the incident edges at the vertices. In (\ref{sws_amp1} the ordering of incident edges 
is chosen separately for each facing pair of dual boundary vertices and the orientations of
the dual boundary edges are induced by the orientations chosen for the \scells{2}.

If $\cg$ is a single loop the choice of orientations for the \scells{2} will determine
orientations for segments of $\cg$ which are generally not
coherent, and the sense of circulation defined by the ordering of incident edges 
at the bivalent vertices will also generally not be coherent. However, the 
amplitudes for incoherently oriented loops can all be expressed in terms
of the amplitude for a coherently oriented loop. Reversing an edge or a bivalent vertex
changes the corresponding spin network function by a factor $(-1)^{2j}$.

The spin networks on the boundaries of 4-cells containing branch lines also have obvious
"canonical" orientation and ordering conventions. Such a spin network consists of two vertices
joined by a collection of edges. The convention requires us to, firstly, pick an ordering of the two 
vertices and orient each edge from the first vertex to the second, and, secondly, to order the
edges the same way at the two vertices. Finally, we may choose the intertwiner bases at the two
vertices to be complex conjugates of each other.

Lets denote the reduced cell amplitudes with these conventions by 
$\bar{\alpha}_\n$. For the empty cells we need to fix no conventions. 
For a 4-cell $\n$ containing a branch line vertex the conventions 
are fixed by the incident branch lines and unbranched components as follows: 
Each edge of the spin network $\cg = S \cap \di\n$ is oriented antiparallel to
the corresponding (overlapping) edges on the boundaries of the neighboring 4-simplices. 
At each vertex of $\cg$ the ordering of the edges is the same as at the facing vertex
on the neighboring 4-cell, and the intertwiner basis is the complex conjugate of that
at the facing vertex.\footnote{These conditions do not fix the orientation and other conventions
uniquely, but the following results are valid whenever the conditions hold.}

When $\bar{w}(S)$ is rewritten in terms of the amplitudes $\alpha_\n$ various sign factors must
be kept track of. Organizing these is not quite trivial. I will simply quote the result
\cite{Rei97x}: Define the "odd surface" $\omega$ to be the surface formed by 
the union of all the unbranched components carrying odd-half-integer spin. 
This surface will generally have only even valence branch lines. Those branch 
lines of $S$ at which only two odd-half-integer spin unbranched components
meet will be taken to define bivalent branch lines of the odd surface. 
Each valence $v$ branch line of $\omega$ contributes a factor $(-1)^{\frac{1}{2} v}$,
each boundary component contributes a $-1$, and finally there is a factor $(-1)^\chi[\omega]$
which is negative for some non-orientable $\omega$.

Our final formula for the reduced spin worldsheet amplitude is therefore
\be             \label{sws_amp2}
\bar{w}(S) = \prod_{\varsigma \mbox{\scriptsize unbranched component}}
          (2j_\varsigma + 1)^{\chi[\varsigma]}\ \prod_{\n\ \mbox{\scriptsize 4-cell of}\ \Pi}
             \bar{\ag}_\n(\cg[S]).
\ee
Notice that the prefactors depend only on the topology of $S$. Only the reduced amplitudes
$\bar{\ag}_\n$ can contain non homeomorphism invariant information.

A simple model, which can be accomodated in our formalism, is Ooguri's lattice formulation of
BF theory \cite{Ooguri}, which in three dimensions is identical with the Ponzano-Regge
model of Euclidean 2+1 GR \cite{PonzReg68}.

The cell amplitude for this model can be written as
\bearr    
a_{\n\,BF}({\bf g}_{\di\n}) & = & \prod_{l} \int dh_l\ v_{\n\,BF}  \label{BF:cellamp} \\
v_{\n\,BF} & = & \prod_{s\ \mbox{\scriptsize wedge of}\ \n} 
\sum_{j_s}\ (2j_s + 1)\: tr\,U^{(j_s)}(g_{\di s}).
\eearr
Here the connection has been extended by defining propagators $h_l$ along edges the $l$ of the 
boundaries of the wedges that connect the center of the 4-simplex with its boundary. The extended
connection defines the holonomies $g_{\di s}$ around the boundaries of the wedges $s$.

Carrying out the integral over $h_l$ one obtains
\be           \label{BF:cellamp2}
a_{\n\,BF}(\bfg_{\di\n}) = \sum_\cg \chi^*_\cg(\mbox{flat}) \chi_\cg(\bfg_{\di\n})
\ee
where $\{\chi_\cg\}$ is a basis of s-net functions on $[\di\n]^*$, and "flat" is a flat
connection on $[\di\n]^*$. $a_{\n\,BF}$ is the flat connection state - integrating any
gauge invariant function of the connection against $a_{\n\,BF}$ yields the value of that
function on flat connections.

From (\ref{BF:cellamp2}) one can read off the reduced amplitudes of cell spin worldsheets:
\begin{itemize}
\item For an empty 4-cell $\bar{\ag}_\n = 1$.

\item For a disk $\bar{\ag}_\n = 1$.

\item For a cell spin worldsheet consisting of a segment of branch 
         line $\bar{\ag}_\n = \dg_{I,I'}$, where $I$ and $I'$ are the intertwiner labels
         at the vertices where the branch line enters and leaves the cell.

\item For a branch line vertex, with cell boundary spin network $\cg = S\cap\di\n$
             $\bar{\ag}_\n = RW[\cg]$, with $RW[\cg]$ the Racah-Wigner recoupling coefficient
                corresponding to the spin network $\cg$.
\end{itemize}

One sees at once from these expressions, and the general formula (\ref{sws_amp2}) for the
reduced spin worldsheet amplitude, that the reduced spin worldsheet amplitude is topological, i.e.
depends only on homeomorphism invariant features of $S$.

\subsection{Geometrical interpretation of spin worldsheets}    \label{geometint}

Spin worldsheets have a natural interpretation as discrete spacetime geometries \cite{Rei94,ReiRov}.
Working in the context of continuum canonical loop quantized GR Rovelli and Smolin  
found that the kinematics implies that the spectrum of the observable measuring the 
area of a given spatial 2-surface $\sg$ is discrete. Any spin network state $|\Cg\rangle$ 
without vertices on the surface or points of tangency to the surface is an eigenstate of 
the area with eigenvalue 
\cite{discarea2}\cite{AL96a}
\be
\mbox{area}_\sg [\Cg] = \mbox{Plank area}\sum_{i \in \sg\cap\Cg} \sqrt{j_i(j_i+1)},
\ee
where $j_i$ is the spin of the spin network at puncture $i$.\footnote{
The action of the area
operator on spin network states with vertices in the surface and points of tangency has since been
found. However, we will not need this action in our lattice context, since the spin networks
we will consider always live on a lattice dual to that of the 2-surfaces whose areas we will 
evaluate. It might also be worth mentioning that the ambiguity in the spectrum of the area
pointed out by Immirzi \cite{Immirzi96a, Immirzi96b, Immirzi97} exists only in the Lorentzian 
theory when loop quantized using the Barbero connection \cite{Barbero94, Barbero95a, Barbero95b}, 
and we are doing Euclidean theory.}
Loll \cite{Loll97} has defined an area operator (${\bf\cal A}_1$ of \cite{Loll97}) in the context
of cubic lattice canonical theory with the same eigenvalues.

Even more directly relevant for us is Ponzano and Regge's model of Euclidean 2+1 GR \cite{PonzReg68}.
This model is just the BF theory described at the ned of \S \ref{integration} in the special case
of a simplicial, three dimensional spacetime $\Dg$. In Ponzano and Regge's original formulation
the connection has been integrated out and the amplitude of a spin network on $\di\Dg$ is given as
a sum over basis spin networks on the boundaries of the 3-simplices, just as in the spin worldsheet
formulation. However, they did not view it as a sum over spin worldsheets but rather as a sum
over spins on the boundaries of the 3-simplices (Intertwiner labels don't figure because the
dual boundaries of 3-simplices have only trivalent vertices, and the space of $SU(2)$ intertwiners for
such vertices is one dimensional). Since the spins must match on neighboring cells, there is really
only one spin on each \scell{2} (or, equivalently, on each 1-cell).

Ponzano and Regge noticed that if one defines the length of each 1-cell to be 
$\mbox{(Plank length)}\times( j + 1/2)$, with $j$ 
the spin on the \scell{2} dual to the 1-cell, then in the large spin limit the amplitude for a 
history in the model is given by the exponential of $i$ times the Regge 
action.\footnote{
Actually it is approximated by the sum of the exponentials of $i$ times the Regge action evaluated 
on a set of different geometries having the same edge lengths, but different deficit angles. 
These alternative geometries are obtained by "folding" the simplicial complex \cite{Barret94}. Like a 
sheet of paper being folded, pressed flat, and then glued to a new sheet of 
paper, the original simplicial complex is mapped continously (by a many to one mapping) to a new 
simplicial complex such that the images of some simplices overlap. The deficit angles are then
computed in the new simplicial complex. }
\footnote{The appearance of the $i$ here is peculiar. One normally uses $\exp (-\mbox{Action})$ as
the weight in Euclidean path integrals. The fact that the $i$ appears might mean that the 
Ponzano-Regge model is not related to the path integral for Lorentzian 2+1 GR as defined by a 
path integral weighted by $\exp (i\mbox{Lorentzian Action})$ (although Barret and Foxon 
\cite{Barret94}
have shown that the Lorentzian classical solutions {\em do} appear, with weight 
$\exp (-\mbox{Lorentzian Action})$, in the Ponzano-Regge state sum. This problem of the $i$ persists
in the Euclidean four dimensional models I will discuss, so it may be that such models are
not directly connected to Lorentzian GR. In any case a quantization of Euclidean GR with the
"wrong complexion", the extra $i$, would provide a very sophisticated toy model.}
The Ponzano-Regge model thus approximates the quantum path integral based on the Regge action
under those circumstances in which the classical approximation is good, i.e. when 1-cell lengths 
are much greater than the Plank length. The classical Regge model in turn approximates continous 
classical gravitational fields down to a resolution given by the largest 1-cell lengths.
Thus the Ponzano-Regge model reproduces classical 2+1 Euclidean GR within its domain of validity,
namely in the classical behaviour of modes of wavelength far above the Planck scale, and
the Ponzano-Regge state sum can be used as a discrete path integral for Euclidean 
2+1 GR. 

The lengths of the 1-cells in $\di\Dg$, the two dimensional "space" of 2+1 GR, 
are determined, according to Ponzano and Regge's geometrical interpretation, by
the spins on the \scells{2} dual to these 1-cells, or, equivalently, by the spins
on the edges of $[\di\Dg]^*$ dual to the 1-cells. Another way to say this is that
the length of a 1-cell in $\di\Dg$ is determined by the spin on the edge of the boundary
spin network that crosses the 1-cell (with the absence of a crossing edge counting as
spin $j = 0$). This was noted by Rovelli \cite{Rov_geom93}, who pointed out that 
Ponzano and Regge's definition of edge lengths as $j + \frac{1}{2}$ times the Planck 
length is equivalent, in the large $j$ limit, to 
that given by the length operator of two dimensional loop quantization, which gives 
$\sqrt{j(j+1)} =  j + \frac{1}{2} + O(\frac{1}{j})$ in Planck units. The length 
operator, when applied to the lattice theory, reproduces the geometry of the metric 
(Regge calculus) formulation of Euclidean 2+1 GR.

Now let's turn to the case of a four dimensional lattice spacetime $\Pi$. On the 
boundary $\di\Pi$, where states live, we will adopt the geometry defined by the 
loop quantized area operator. That is, in a spin network basis state, corresponding to
the spin network $\Cg$ on $\di\Pi$ we define the area of a 2-cell in $\di\Pi$ to be 
$\sqrt{j(j+1)}$ times the Planck area, with $j$ the spin of the edge of $\Cg$, 
if any, that punctures the 2-cell (again the absence of such an edge counts as 
$j = 0$). 

This geometrical interpretation of spin networks on the boundary can be 
extended to a geometrical interpretation of spin worldsheets in spacetime. Fix a 
2-cell $\sg$ in $\Pi$. It is always possible
to define a three dimensional cellular complex, $\Sg$, containing $\sg$ that divides $\Pi$ into two 
halves. (For instance, one can take $\Sg$ to be
the boundary of a 4-cell that $\sg$ belongs to). The spin network formed by the intersection of a 
spin worldsheet $S$ and $\Sg$ defines an intermediate state in the history of the gravitational 
field represented by the spin worldsheet. The area of $\sg$ is thus given, via the same expression 
that applies on $\di\Pi$, by the spin of the edge of $S\cap\Sg$ that punctures $\sg$. This is just
the spin of $S$ where it punctures $\sg$. 
The geometrical interpretation of a spin worldsheet, $S$, is therefore the following \cite{Rei94,
ReiRov}: Each 2-cell $\sg$ of $\Pi$ that is punctured by $S$ has non-zero area, given by
\be          \label{spinws_area}
\mbox{area}\sg[S] = \mbox{Planck area}\times\sqrt{j(j+1)},
\ee
with $j$ the spin on $S$ where it punctures the 2-cell. 2-cells not punctured by the spin worldsheet
have area zero. 

In this way the sum over spin worldsheets can be interpreted as a sum over discrete 
spacetime geometries, which is of 
course interesting, since the Planck scale discreteness might provide a cure for the ultraviolet
divergences of both GR and theories of matter living in the geometrical background established by
the gravitational field. Note that the geometrical interpretation of spin worldsheets given here
is also well defined for spin worldsheets living in the continuum. (\ref{spinws_area})
then applies to any 2-surface $\sg$ puncturing the spin worldsheet only once (transversely at an
interior point of an unbranched component).
The argument used to derive the geometrical interpretation from the canonical kinematics also 
works just as well in the continuum.

Another way to state the geometrical interpretation of the spin worldsheet, which shows how
very closely analogous it is to Ponzano and Regge's interpretation of their model is to say
that the area of a 2-cell in $\Pi$ is determined by the spin on its dual \scell{2} according to
(\ref{spinws_area}).

The same type of argument can be used to assign volumes\footnote{ 
Volumes are {\em a priori} 
independent of the areas since no particular internal geometry of the cells is assumed.}
to 3-cells punctured by branch lines,
which are functions of the intertwiner carried by the branch line at the point of 
puncture \cite{ReiRov}. However I will make no use of this possible geometrical interpretation of the intertwiner 
assignments in the present work.

There is a caveat that the reader should keep in mind. The canonical area operator that
has been used as the basis of our geometrical interpretation of spin worldsheets is
obtained by quantizing a classical measure of the area by simply substituting operators
for the canonical variables in the classical formula. It has {\em not} been shown that 
the area of a surface obtained via this operator really reduces to the classical area in
a classical limit. Similarly, we can't be {\em sure} that the spacetime geometry we have defined
for worldsheets really reproduces classical spacetime geometry in the classical limit.
On the other hand, Geometry is a mathematical construct that we may define any way that is 
convenient to us, and the geometry we have defined is sufficiently simple and elegant that
it may well be a useful concept even if it does not reduce to the classical geometry in the classical
limit.
Furthermore, it is encourageing that the analogous geometrical interpretation of the 
Ponzano-Regge state sum does reproduce classical geometry in the classical limit, and I will adopt the geometrical 
interpretation (\ref{spinws_area}) as a guide in searching for a good dynamical model.

\section{A proposal for a model of Euclidean quantum GR} \label{model}

\subsection{The model}               \label{simplicial model}

In this section I propose a specific model for Euclidean quantum GR. It is based on Plebanski's
form of GR \cite{Plebanski77}\cite{CDJM}. More specifically, it is a lattice version of the path integral
quantization with histories weighted by $\exp (i\:\mbox{Euclidean action})$.
As already pointed out in the context of the Ponzano-Regge model, such a quantization of the
Euclidean theory is rather unconventional, and its relation to the physical Lorentzian theory 
(in which, at least in the semiclassical sector, the weights
of histories should be $\exp (i\:\mbox{Lorentzian action})$. Nevertheless Euclidean GR quantized
in this way is interesting, at the very least, as a toy model.


Spacetime is represented by an orientable four dimensional simplicial complex $\Dg$. The derived 
complex 
$\Dg^+$, with 4-cells formed by the intersections of the 4-simplices of $\Dg$ and the 4-cells
of its duel complex $\Dg^*$, plays an essential role in the definition of the model. In particular 
the wedges are crucial. Recall that a wedge $s(\sg,\n)$ is a 2-cell of $\Dg^+$ formed by
the intersection of a 4-simplex $\n$ and a 2-cell $\sg^*$ of $\Dg^*$ dual to a 2-simplex $\sg$ in
the boundary of $\n$.
(See \S \ref{cell:structure} for a discussion of the derived complex, and Fig. \ref{wedges} for an 
illustration of a wedge).

In the model a spacetime field configuration (or history) is specified by a collection of spins 
and $SU(2)$ group elements (parallel propagators). Each wedge $s$ carries a spin 
$j_s \in \{0, \frac{1}{2}, 1, \frac{3}{2}, 2, ...\}$. The 1-cells of $\Dg^+$ that bound wedges
each carry an $SU(2)$ parallel propagator, defining a lattice connection that specifies the 
holonomy around the boundary of each wedge. 

The amplitude of such a history is
\be           \label{model_def}
w = \prod_{\n\ \mbox{\scriptsize 4-cell of}\ \Dg}   tr^{j_1\otimes...\otimes j_{10}}
[\ e^{-\frac{1}{2z^2}\Omega_{ij} \Omega^{ij}}\ \bigotimes_{s\ \mbox{\scriptsize wedge of}\ \n} 
(2j_s + 1) U^{(j_s)}(g_{\di s})].
\ee
Each 4-simplex $\n$ contributes a factor, a trace in a finite dimensional Hilbert space associated
with the 4-simplex. The $g_{\di s}$ are the holonomies around the boundaries of the wedges $s$, with 
base point at the center of $\n$, and the 
$U^{(j_s)}(g_{\di s})$ are the spin $j_s$ representation matrices of these holonomies, each 
realized on a separate spin $j_s$ representation carrying space associated with to its wedge $s$.

The trace is taken in the tensor product ${\cal H}_\n^{j_1\otimes...\otimes j_{10}}$ of these $10$ 
carrying spaces, as indicated by the superscript $j_1\otimes...\otimes j_{10}$. 
Finally, $z \in \Real$ is a constant adjustable 
parameter of the model\footnote{
Note that if we take $z \rightarrow 0$ then
\be
e^{-\frac{1}{2z^2}\Omega_{ij} \Omega^{ij}} = P_{ker\Omega},
\ee
with $P_{ker\Omega}$ the orthogonal projector onto $ker\Omega$, the intersection of the kernels of 
the five independent components of $\Omega_{ij}$. In the original model proposed in \cite{Rei96}
this projector appeared in place of the gaussian. However, while $z = 0$ is a natural choice,
I will argue that it may lead to difficulties.}
, and the operator $\Omega_{ij}$ is defined as 
follows: Number the vertices of $\n$ 1,2,3,4,5, so that the vectors 12, 13, 14, 15 
form a basis with the same orientation as $\n$, then
\bearr
\Omega_{ij} & = & \Cg_{ij} - \frac{1}{3}\dg_{ij} \Cg_k{}^k\label{Omega_operator}\\
\Cg_{ij} & = & \frac{1}{4}\sum_{P,Q,R,S,T  \in \{1,2,3,4,5\}} J_{PQR\,i}\otimes J_{PST\,j}
\eg^{PQRST}.        \label{Gamma_operator}
\eearr
($i\,,j\,\in\{1,2,3\}$). That is, $\Omega$ is the trace free part of $\Cg$. In (\ref{Gamma_operator})
each oriented 2-simplex $\sg$ of $\n$ is represented by the triplet of its vertices ordered in a 
positive sense around its boundary. $J_{PQR\,i}$ is the vector of $SU(2)$ generators 
acting on the holonomy $g_{\di s(PQR,\n)}$ around the wedge $s(PQR,\n)$ associated 
with the 2-simplex $PQR$.

Finally, $\eg^{PQRST}$ is the antisymmetric symbol with $\eg^{12345} = 1$.
Note that since $\Omega_{ij}$ is traceless and symmetric in $ij$, it is a spin $2$ tensor operator
acting in ${\cal H}_\n^{j_1\otimes...\otimes j_{10}}$. 

$\Gamma_{ij}$ can be written more compactly as
\be
\Gamma_{ij} = \sum_{s,\bar{s}\ \mbox{\scriptsize wedges of}\ \n}  J_{s\,i}\otimes J_{\bar{s}\,j}
sgn(s,\bar{s}).   \label{Gamma_operator2}
\ee
For each pair of wedges $(s,\bar{s}) = (s(PQR), s(PST))$ of $\n$ $sgn(s,\bar{s}) = \eg^{PQRST}$.
$sgn(s,\bar{s})$ can also be defined more geometrically as the sign of the 4-volume spanned by
by the 2-simplices $PQR$ and $PST$, i.e. by the ordered set of vectors $\{PQ,\:PR,\:PS,\:PT\}$,
with $sgn(s,\bar{s}) = 0$ when the volume is zero. In fact, in any linear coordinate system $x^\ag$
on $\n$ respecting the orientation of $\n$,
\bearr
sgn(s,\bar{s}) & = & \frac{1}{4! V_\n} t_{PQR}^{\ag\bg} t_{PST}^{\cg\dg} \eg_{\ag\bg\cg\dg}
\label{simp_wedge_simp}\\
                          & = & \frac{25}{16 V_\n} t_{s}^{\ag\bg} t_{\bar{s}}^{\cg\dg} \eg_{\ag\bg\cg\dg},
\label{wedge_wedge_wedge}
\eearr
where $V_\n$ is the coordinate 4-volume of $\n$, for any 2-cell $c$, 
$t_c^{\ag\bg} = \int dx^\ag\wedge dx^\bg$ is the coordinate area bivector of $c$. $sgn$ is thus
simply the translation into simplicial terms of the spacetime antisymmetric tensor density
$\eg_{\ag\bg\cg\dg}$.\footnote{
The definition of exterior multiplication used here is
$[a \wedge b]_{\ag_1 ... \ag_m\bg_1 ... \bg_n} = a_{[\ag_1 ... \ag_m}
b_{\bg_1 ... \bg_n]}$, where spacetime indices are labeled by lower
case greek letters $\{\ag,\bg,\cg,...\}$. Forms are integrated according 
to $\int_A a
= \int_A \eg^{u_1 ... u_m} a_{u_1 ... u_m}\ d^m\sg$ where $A$ is 
an $m$ dimensional manifold, $\sg^u$ are coordinates on $A$, the indices
$u_i$ run from $1$ to $m$, and $\eg^{u_1 ... u_m}$ is the $m$ dimensional
Levi-Civita symbol ($\eg^{12...m} = 1$ and $\eg$ is totally antisymmetric).}

The model defined by (\ref{model_def}) fits perfectly into the framework of \S \ref{lattice}.
The cell (4-simplex) amplitude $a_\n$ as a function of only the connection ${\bf g}_{\di\n}$ 
on the cell boundary, is obtained by taking the factor contributed by that cell in (\ref{model_def}),
and summing over the spins and integrating over the connection in the interior of the cell.
Denote by $h_l$ the parallel propagators along the edges $l$ in the boundaries of wedges that 
connect the center of the 4-simplex with its boundary, then the cell amplitude is
\be          \label{model:cellamp}
a_\n({\bf g}_{\di\n}) = \prod_{l} \int dh_l\ \prod_s \sum_{j_s}\  tr^{j_1\otimes...\otimes j_{10}}
[e^{-\frac{1}{2z^2}\Omega_{ij} \Omega^{ij}}\ \bigotimes_{s} (2j_s + 1) U^{(j_s)}(g_{\di s})].
\ee

\subsection{Motivation of the model}     \label{motivation}

Why should this model be a quanization of GR? If one replaces the operator 
$\hat{C} = exp(-\frac{1}{2z^2}\Omega_{ij} \Omega^{ij})$ with $\One$ in the trace in (\ref{model_def}) one 
obtains Ooguri's simplicial formulation of $SU(2)$ BF theory \cite{Ooguri} (see (\ref{BF:cellamp})):
\be           \label{Ooguri_BF}
w_{BF} = \prod_{s\ \mbox{wedge of}\ \Dg^+} (2j_s + 1)\: tr\,U^{(j_s)}(g_{\di s}).
\ee
$SU(2)$ BF theory, which in the continuum formulation has the classical action 
$\int \Sg_i \wedge F^i$,\footnote{
The field $\Sg$ is usually called "$B$", hence the name "BF theory". Here we use $\Sg$
to be consistent with the notation of \cite{CDJ} for the Plebanski model.}
with $\Sg_{i\,\ag\bg}$ a triplet of 2-form fields and $F^i$ the curvature of an $SU(2)$ 
connection, is a topological field theory which
is closely related to GR. If one constrains $\Sg$ to satisfy the "metricity constraint" 
\be               \label{metricity_Pleb}
\Sg_i\wedge \Sg_j - 1/3 \Sg_k\wedge \Sg^k = 0
\ee
one obtains Plebanski's formulation of full GR \cite{Plebanski77}\cite{CDJ}\cite{Rei95}. 
When (\ref{metricity_Pleb}) holds and $\Sg$ is non-degenerate in the sense that 
$\Sg_k\wedge\Sg^k \neq 0$ one can construct a metric out of the  $\Sg_i$. The metrics corresponding 
in this way to non-degenerate solutions of Plebanski's theory are exactly the solutions to 
Einsteins field equations.\footnote{
Plebanski's theory also has degenerate solutions, in which the metric is degenerate or altogether
undefined, though some geometrical quantites such as areas of surfaces are still defined. It is an 
extension of standard GR to geometries on which
Einsteins field equations are not well defined. Because this extension is
not unique (the Samuel-Jacobson-Smolin action \cite{Samuel87}\cite{Jacobson_Smolin88a} defines
a distinct extension)\cite{Rei95}, and because the degenerate sector may be important 
for the quantum theory, I refer Plebanski's formulation of GR as Plebanski's theory.} 

The insertion of the operator $\hat{C}$ in (\ref{model_def}), which supresses states in 
${\cal H}^{j_1\otimes ...\otimes j_10}_\n$ 
with large expectation values of $\Omega_{ij}$ in (\ref{model_def}) is supposed to be the 
counter part of the metricity constraint in the classical continuum theory.

To explain this more fully let me make a detour and present the properly Plebanski theory, 
and also a variant of the classical simplicial model of \cite{Rei97a}, that reproduces the Plebanski 
theory in the continuum limit. The Plebanski theory is defined by the action
\begin{equation}        \label{Plebanski_action}
I_P = \int \Sg_i \wedge F^i - \frac{1}{2} \phi^{ij} \Sg_i \wedge \Sg_j
\end{equation}
(The euclidean theory is obtained when all fields are real). The lagrange multiplier
$\phi^{ij}$ is a traceless symmetric matrix. The stationarity of the action with respect to
this field requires precisely the metricity condition (\ref{metricity_Pleb}). The action is invariant
under $SU(2)$ gauge transformations and diffeomrphisms. The indices, which run over $\{1,2,3\}$
are spin 1 (adjoint) vector indices under the $SU(2)$ gauge transformations.

On non-degenerate solutions the fields $\Sg$, $\phi$, and the $SU(2)$ connection $A$ can be 
expressed in terms of more conventional variables. $\Sg$ is the self-dual part of the
vierbein wedged with itself:\footnote{
The adjoint representation of $SU(2)$ is the fundamental of $SO(3)$, so upstairs and 
downstairs adjoint representation indices are the same.}
\begin{equation}        \label{Sigma_e_wedge_e}
\Sg_i = 2[e \wedge e]^{+\,0i} \equiv e^0 \wedge e^i + \frac{1}{2}
        \eg_{ijk} e^j \wedge e^k,
\end{equation}
which transforms as a spin 1 vector under $SU(2)_L$, the left-handed 
subgroup of
the frame rotation group $SO(4) = SU(2)_R \otimes SU(2)_L$, 
and as a scalar under $SU(2)_R$. $A$ is the 
self-dual ($SU(2)_L$) part of the spin connection, and $\phi$ turns out to be
the left-handed Weyl curvature spinor.

Ashtekar's canonical variables are just the purely spatial parts of
$A$ and $\Sg$ (the dual of the spatial part of $\Sg$ is the densitized triad),
and, in the non-degenerate sector, the canonical theory derived from 
(\ref{Plebanski_action}) is identical to Ashtekar's \cite{CDJM}\cite{Rei95}. 
Since this is precisely the sector of non-degenerate spatial metric it is of course
also equivalent to the ADM theory \cite{ADM62}. However, when the metric is 
degenerate the canonical theory differs from Ashtekar's \cite{Rei95}.

A classical simplicial model, which reduces in the continuum limit to the 
Plebanski theory, is given in \cite{Rei97a}. We will make use of a slightly modified
variant of that model, which has the same continuum limit. The fundamental variables of this 
model are, in addition to the lattice connection, defined as for the quantum model 
(\ref{model_def}), an $SU(2)$ spin 1 vector $e_{s\,i}$ associated to each wedge $s$, 
which will more or less play the role of Plebanski's $\Sg_i$ field,\footnote{%
We define $e_{\sg\n}$ to reverse sign when the orientation of $\sg$ is 
reversed.}
and a spin 2 $SU(2)$ tensor $\varphi_\n$ (represented by a symmetric, 
traceless matrix, $\varphi^{ij}_\n$) associated with each 4-simplex. 
$\varphi^{ij}_\n$ plays the role of $\phi^{ij}$.  

The action for the model is
\be      \label{simplicial_action}
I_{\Dg}^o = \sum_{\n\ \mbox{\scriptsize 4-simplex of}\ \Dg}[\  
\sum_{s\ \mbox{\scriptsize wedge of}\ \n} e_{s\,i}\rho^i_s - \frac{1}{60}
\varphi^{ij}_\n \sum_{s,\bar{s}\ \mbox{\scriptsize wedges of}\ \n} 
e_{s\,i}e_{\bar{s}\,j} sgn(s,\bar{s})\ ].  
\ee
$\rho^i_s$ is a measure of the curvature on $s$. It is a function of the 
$SU(2)$ parallel propagators via
\be
g_{\di s}  =  e^{i\rho_s\cdot J} = \cos\frac{|\rho_s|}{2}\One 
+ 2i\sin\frac{|\rho_s|}{2} \hat{\rho_s}\cdot J,                     \label{rho_def}
\ee
where $g_{\di s}$ the holonomy around $\di s$, the boundary of the wedge, 
$\hat{\rho}$ is the unit vector $\rho/|\rho|$, and the 
$J_i$ are $1/2$ the Pauli sigma matrices\footnote{%
\be
\begin{array}{ccc}
J_1 = \frac{1}{2}\left[ \begin{array}{cc} 0 & 1 \\ 1 & 0 \end{array}\right] &
J_2 = \frac{1}{2}\left[ \begin{array}{cc} 0 & -i \\ i & 0 \end{array}\right] &
J_3 = \frac{1}{2}\left[ \begin{array}{cc} 1 & 0 \\ 0 & -1 \end{array}\right].
\end{array}
\ee}
$\rho_s$ might also be called the ''rotation vector" because it is the vectorial angle
of the rotation produced by the holonomy $g_{\di s}$.
\footnote{%
Note that $\rho_s$ reverses sign when the orientation of $s$ reverses because the 
direction of the boundary $\di s$ reverses.}
\footnote{
$\rho_s^i$ is singular at $g_{\di s} = -\One$, but this does not cause any problem in the
classical theory.}


To recover the Plebanski theory in the continuum limit one must identify
the simplicial complex with the spacetime manifold\footnote{%
The sequence of ever finer simplicial decompositions of the spacetime manifold used to
define the continuum limit has to satisfy certain "fatness" conditions ensuring that
the piecewise linear structure of the simplicial complexes, and their derived complexes,
are compatible with the differentiable structure of the manifold.}
, the lattice
connection with the parallel propagators along the lattice edges computed from the
connection $A$, $e_{s(\sg,\n)\,i}$ with the integral $\int_\sg \Sg_i$\footnote{
The integrand $\Sg_i$ is parallel transported from its home on $\sg$, along a 
straight line (according to the linear structure of $\n$, first to the center of 
$\sg$ and from there to the center of $\n$, before the integral is taken. This way 
the integral is a well defined $SU(2)$ vector living at the center of $\n$. In 
\cite{Rei97a} a more elaborate definition of $e_{s\,i}[\Sg]$ is used to make proofs 
cleaner, but the definition given here is adequate.}
of $\Sg_i$ over the
2-simplex $\sg$, and $\varphi_\n^{ij}$ with the value of $\phi^{ij}$ at the center of 
the 4-simplex $\n$.

Stationarity of the action (\ref{simplicial_action}) with respect to variations of 
$\varphi_\n^{ij}$ requires a lattice version of the metricity condition 
(\ref{metricity_Pleb}) to hold. Namely, it requires
\be          \label{metricity_simp}
\omega_{ij} = \cg_{ij} - \frac{1}{3}\dg_{ij} \cg_k{}^k = 0,
\ee
where
\be
\gamma_{ij} = \sum_{s,\bar{s}\ \mbox{\scriptsize wedges of}\ \n} 
e_{s\,i}e_{\bar{s}\,j} sgn(s,\bar{s}).
\ee

Indeed (\ref{metricity_simp}), a non-degeneracy condition, $\cg_k{}^k \neq 0$, and the 
identification of $e_{s(\sg,\n)\,i}$ with the integral of $\Sg_i$ over the 2-simplex $\sg$
imply that in the continuum limit, and on flat solutions, $e_{s(\sg,\n)\,i}$ is the {\em metric
normal area vector of $\sg$} in an orthonormal reference frame such that $\sg$ is purely spatial. 
(see appendix A of \cite{Rei97a}). This metrical significance of the $e_s$ can also be understood 
from the point of view of the Ashtekar variables: With the integrands parallel transported
to the center of $\sg$ along straight lines (according to the linear structure of $\sg$)
\be 
\int_\sg \Sg_i = \int_\sg \tilde{E}^a_i dn_a
\ee
where the dreibein density $\tilde{E}$ and and the coordinate normal area element $dn^a$ are
evaluated in spacetime coordinates for which the cell boundary $\di\n$ is an equal time 
hypersurface.

Now, let's return to the motivation of the quantum model (\ref{model_def}).  
It will not {\em proven} in any sense that the quantum model (\ref{model_def}) really reproduces
Euclidean GR as its classical limit. However, it will be shown how it can be obtained
as a formal quantization of the classical simplicial model defined by $I^o_\Dg$. 

Since, this
classical model has a finite set of degrees of freedom for a finite simplicial complex
straightforward path integral methods yield a true quantization of this model:
The boundary data for $I^o_\Dg$ is the connection. Thus one can define a quantization
by setting the amplitude for a given boundary connection to be the integral of 
$\exp (i I^o_\Dg)$\footnote{
The Planck area $a_{\scriptsize \mbox{Plank}} = G\hbar/ c^3$, which divides the action in the
exponential, is set to $1$.}
over all histories matching the given connection on $\di\Dg$. This exact path 
integral quantization has not been completed, but it is sure to lead to a model 
that is more complicated than (\ref{model_def}). (It is noteworthy,
however, that preliminary results indicate that the spin 2 tensor operator 
$\Omega_{ij}$ that is so central to (\ref{model_def}) also plays a central role
in the exact path integral quantization). 

The advantage of an exact path integral quantization based on the action $I^o_\Dg$ is that its
naive classical limit, in which only histories near an extremum of the action contribute to 
the path integral, reproduces classical Euclidean GR.
One would expect the naive 
classical limit to be realized if the simplicial complex $\Dg$ is chosen coarse enough that, 
according to the classical solution correponding to the given boundary data, 
the average 4-volume per cell is much larger than the Planck 4-volume. Then the path integral
should be dominated by histories in which most 4-simplices are much larger than the Planck scale,
in the sense that their 2-simplices have areas much larger than the Planck area. Such large
4-simplices are each individually in the classical regime, so a semi-classical evaluation of the 
amplitude for the simplex, dominated by the near extrema of the action, should be accurate.
As a result the effective classical action is, modulo small corrections, the "bare" action used 
to define the path integral. 

However, the classical limit of a continuum quantum model defined
as some sort of limit of simplicial models as the simplicial complex becomes infinitely
fine, then the relation between the cell amplitudes and the classical
action is less direct. When $\Dg$ is taken fine enough that the classical
solution assigns an average 4-volume to the 4-simplices that is much smaller than the Planck 
4-volume then one would expect that the path integral is dominated by histories in which the
2-simplices have areas that are of the order of the Planck scale, or zero. Each 4-simplex
individually is subject to large quantum fluctuations, and there is no particular reason
to expect the effective classical action for the complex to be well approximated by the bare action.

Thus the cell amplitude obtained from an exact path integral quantization based on $I^o_\Dg$
may not be a good choice for models living on very fine complexes. A better choice as a first guess
may be an amplitude, like that defined by (\ref{model:cellamp}), that is inspired by the cell 
amplitude defined by $I^o_\Dg$, but is simpler.


The cell amplitude for the model obtained by path integral quantization with $I^o_\Dg$ is
\bearr           
a_{\n\,o}(\bfg_{\di\n}) & = & \prod_l \int_{SU(2)} dh_l\ v_{\n\,o}(\{g_{\di s}\})  
\label{int_amplitude} \\
v_{\n\,o}(\{g_{\di s}\}) & = & \prod_s (\frac{1}{2\pi})^3\int d^3 e_s \ (\frac{1}{120\pi})^5\int d^5\varphi_\n 
              \ e^{i \sum_{s} e_{s\,i}\rho^i_s - \frac{1}{60}
         \varphi^{ij}_\n \sum_{s,\bar{s}} 
               e_{s\,i}e_{\bar{s}\,j} sgn(s,\bar{s})},         \label{hol_amplitude}  \\
       & = & \prod_s (\frac{1}{2\pi})^3 \int d^3 e_s\ \dg^5(\omega(e))\ e^{i \sum_{s} e_{s\,i}\rho^i_s},
\eearr
where the $h_l \in SU(2)$ are the parallel propagators along the edges $l$ (of the wedges) 
connecting the center of $\n$ and the centers of it's 3-simplex faces, and $\dg^5$ is the 
delta distribution on traceless symmetric $3\times 3$ matrices.\footnote{
$\dg^5(X) = 
2\sqrt{3}\ \dg(X_{12})\dg(X_{23})\dg(X_{31})\dg(X_{11} - X_{22} + X_{33})\dg(X_{11} - X_{33})$.}
(The constant numerical factors in the definition (\ref{hol_amplitude}) of the unnormalized
amplitude $v_{\n\,o}$ have been chosen to simplify later results).

$v_{\n\,o}$ can be thought of as the wavefunction of a particle in $(\Real^3)^{10}$, 
with coordinates $\rho^i_s$.\footnote{
Only $\rho^i_s$ such that $|\rho_s| \leq 2\pi$ correspond to holonomies 
$g_{\di s}$.}
In particular, it is the state obtained by acting on the wavefunction 
\be
\Psi = \prod_s (\frac{1}{2\pi})^3 \int d^3 e_s\ e^{i \sum_{s} e_{s\,i}\rho^i_s} 
              = \prod_s \dg^3(\rho_s)
\ee
with the operator 
\be
\hat{C}_o = \dg^5(\omega(\hat{e})),
\ee
where $\hat{e}_{s\,i} = -i \frac{\di}{\di \rho^i_s}$ is the momentum conjugate to $\rho_s^i$.

Now, the delta distribution state $\Psi$ can be written as
\be
\Psi = (2 \pi^2)^{10}\prod_{s\ \mbox{\scriptsize wedge of}\ \n} 
                         \sum_{j_s} (2j_s + 1)\: tr\,U^{(j_s)}(g_{\di s})
\ee
in the domain $|\rho| \leq 2\pi$.
Notice that (aside from the unimportant numerical factor) this is just the amplitude $v_{\n,BF}$
for the holonomies $g_{\di s}$ in $\n$ in BF theory, obtained by summing the amplitude 
for a cell given in (\ref{Ooguri_BF}) over spins.
In other words, $v_{\n\,o}$ is obtained by acting on the corresponding BF theory amplitude 
$v_{\n,BF}$ with $\hat{C}_o$ which takes out the "non-metrical modes" - the part of
$v_{\n,BF}$ orthogonal to the intersection of the kernels of the operators $\omega_{ij}(\hat{e})$.

The metricity constraint can be softened somewhat by replacing $\dg^5(\omega(\hat{e}))$ by
a gaussian 
\be
\hat{C}_{o\,z} = e^{-\frac{1}{2z^2}\omega_{ij}\omega^{ij}}.
\ee
(The factor $\frac{1}{\sqrt{2\pi} z}$ required to normalize the gaussian is absorbed
in the over all normalization of the sum over histories).

The amplitude for wedge holonomies, $v_\n$, of (\ref{model_def}) is obtained by replacing
$\hat{e}_{s\,i}$ by the generator $J_{s\,i}$ in the operator $\hat{C}_{o\,z}$ and acting
with the resulting operator
\be
\hat{C} = e^{-\frac{1}{2z^2}\Omega_{ij}\Omega^{ij}}
\ee
on $\Psi$. 

$\hat{e}_{s\,i} = J_{s\,i}$ at $g_{\di s} = \One$, so these operators act the same
way on $\Psi$, as do $\omega_{ij}(\hat{e})$ and $\Omega_{ij}$ since these are linear
in each $\hat{e}_s$ and $J_s$ respectively. Polynomials in $\hat{e}_s$ and $J_s$ act
differently since the $J_{s\,i}$ don't commute while the $\hat{e}_{s\,i}$ do. However,
in the naive classical limit the path integral is dominated by histories in which 
the areas of the 2-simplices are much larger than the Planck area. That is to say,
$|e_{s(\sg,\n)}| \gg 1$. In this limit the difference between low order polynomials $P(\hat{e})$
and $P(J)$ is sub-leading order. If had kept conventional units instead of setting
$a_{Planck} = 1$ then the difference would consist of order $a_{Planck}$ terms. 
Unfortunately this sort of argument does not work for infinite power series like
that defining $\hat{C}$, so there is no proof that $\hat{C}\Psi$ approximates
$\hat{C}_{o\,z}\Psi$.

The model (\ref{model_def}) is thus a formal quantization of the classical simplicial model
with action $I^o_\Dg$.

It {\em is} possible to prove that $\hat{C}$ inforces the metricity constraint for
nearly continous histories, that is histories for which the total curvature, $\rho^i_s$, 
on each wedge is very small. In this limit
\be
\omega_{ij}(\hat{e}) \hat{C} \Psi \simeq \Omega_{ij} \hat{C} \Psi.
\ee
$\Omega_{ij}\hat{C} = \Omega_{ij}e^{-\frac{1}{2z^2}\Omega_{kl}\Omega^{kl}}$ is a hermitian
operator with eigenvalues of absolute value $\leq z$. It follows that if $z$ is taken very 
small violations of metricity will be very small.

%
%
%

Is the substitution $\hat{e} \rightarrow J$ consistent with the geometrical interpretation
of spin worldsheets given in \S \ref{geometint}? It seems to be. 

Recall that in the classical simplicial model defined by $I^o_\Dg$ the area of a 2-simplex $\sg$,
in a nearly continous history which is near a continuum limit solution, is approximately
$|e_{s(\sg,\n)}|$, where $\n$ is any 4-simplex incident on $\sg$. The field equations ensure
that $|e_{s(\sg,\n)}|$ is essentially independent of the choice of $\n$ on such histories.
In the path integral quantization of this model it is thus consistent with the naive classical 
limit to take $|e_{s(\sg,\n)}|$ as the area of $\sg$ estimated within $\n$, with the true
area of $\sg$ defined as the common value of the $|e_{s(\sg,\n)}|$ when they agree.

Now, recall that $v_{\n\,o}(\{g_{\di s}\}) = \hat{C} \Psi$ is the amplitude for the holonomies
$g_{\di s}$ in $\n$, obtained by integrating the amplitude $exp(i\ \mbox{action of $\n$})$ over
the $e_{s\,i}$ living in $\n$. The amplitude for {\em both} the holonomies in $\n$ and a given
value of $e_{s\,i}$ is the component with eigenvalue $e_{s\,i}$ in an expansion of $v_{\n\,o}$
into eigenfunctions of the operator $\hat{e}_{s\,i}$. Similarly, the amplitude for the holonomies
and given values of the areas $|e_{s}|$ in $\n$ is obtained by expanding $v_{\n\,o}$ into
eigenfunctions of the area operators $\sqrt{\hat{e}_s^2}$.

I will now argue that the expansion 
\be        \label{delta_expansion}
\Psi = (2 \pi^2)^{10}\prod_{s\ \mbox{\scriptsize wedge of}\ \n} \sum_{j_s} (2j_s + 1)\: tr\,U^{(j_s)}(g_{\di s})
\ee
produces essentially such an expansion of $v_{\n\,o}$ into area eigenfunctions.
$\sqrt{\hat{e}_s^2}$ commutes with $\hat{C}_o$ and\footnote{
{\em Proof}: 
\be
\hat{e}^2 = -\frac{\di}{\di \rho^i}\frac{\di}{\di \rho_i}
\ee
and
\bearr
\hat{e}^2 tr U^{(j)} & = & - [\frac{d^2}{d|\rho|^2} 
+ \frac{2}{|\rho|}\frac{d}{d|\rho|}]
\frac{\sin (j+\frac{1}{2})|\rho|}{\sin \frac{1}{2} |\rho|} \\
 & = & j(j+1) tr U^{(j)}    \\
 & & \ \ \ + (\frac{\cos \frac{1}{2}|\rho|}{\sin \frac{1}{2}|\rho|} 
  - \frac{2}{r})[(j + \frac{1}{2})
\frac{\cos (j + \frac{1}{2})|\rho|}{\sin \frac{1}{2}|\rho|} - \frac{1}{2}
\frac{\sin (j+\frac{1}{2})|\rho|\ \cos\frac{1}{2}|\rho|}{\sin^2 \frac{1}{2} 
|\rho|}].
\eearr
The result then follows trivially when it is noted that the expression in 
the last line above is bounded on $SU(2)$.}
\be
\lim_{j_s \rightarrow \infty} \frac{1}{j_s (j_s + 1)} \hat{e}_s^2 tr U^{(j_s)} = tr U^{(j_s)},
\ee
so the terms $(2 \pi^2)^{10} \hat{C}_o\:\prod_{s\ \mbox{\scriptsize wedge of}\ \n} 
(2j_s + 1)\: tr\,U^{(j_s)}$ are approximate eigenfunctions of the areas with eigenvalues  
$|e_s| = \sqrt{j_s(j_s +1)}$ in the limit of large $j_s$, and thus large areas. It is of course 
only in this limit, of $|e_s| \gg 1 = \mbox{Planck area}$, that the naive classical limit
which justifies our interpretation of $|e_s|$ as area is expected to hold. Identifying
$j_s(j_s + 1)$ with the area in Planck units is thus quite consistent with the requirement that
the path integral quantization of $I^o_\Dg$ reproduces classical GR when the naive classical limit 
is valid.

The model (\ref{model_def}) is obtained by the modification $\hat{C}_o \rightarrow \hat{C}_{o\,z}$,
which is irrelevant for the present discussion, and the substitution $\hat{e}_s \rightarrow J_s$.
We shall see in \S \ref{model_sws} that in a spin worldsheet formulation of that model the spins 
$j_s$ turn out to be exactly the spins of the worldsheets: Each spin distribution that has non-zero 
amplitude, once the connection is integrated out, is represented by spin worldsheets with the same
distribution of spins on the wedges. The substitution $\hat{e}_s \rightarrow J_s$ leads to
the geometrical interpretation of spin worldsheets of \S \ref{geometint}.

The reader might wonder how the discrete spectrum of areas defined by the spins arose from the
continous spectrum of $\sqrt{\hat{e}_s^2}$? The key step is (\ref{delta_expansion}). The right
side provides an expansion of the delta distribution $\Psi$ on the $\rho_s$ in the compact
domain $|\rho_s| \leq 2\pi\ \forall s \mbox{in}\ \n$ corresponding to the product of group 
manifolds $SU(2)^{10}$. Instead of requiering the entire continuum of $\sqrt{\hat{e}_s^2}$ 
eigenfunctions, this requires only a countable set of (approximate) eigenfunctions, with the
discrete spectrum of eigenvalues defined by the spins. The situation is quite analogous to what
happens when a function is Fourier expanded on a finite interval instead of the whole real line.

\subsection{Spin worldsheet formulation of the model}        \label{model_sws}

The cell amplitude (\ref{model:cellamp}) is very easily expanded on a spin network basis.
It is already given as a sum over spins on the wedges, so let's focus on a single term in
this expansion, with a particular assignment $\{j_s\}$ of spins to the wedges. 
Each of the edges $l = C_\n C_\tau$ connecting the center of the 4-simplex $\n$ with one of
its 3-simplex faces $\tau$ is shared by four wedges. The $h_l$ dependent part of 
$\bigotimes_s U^{(j_s)}(g_{\di s})$ is therefore a direct product of four representation
matrices of $h_l$. Integrating over $h_l$ replaces this product with the projector
$\sum_I W^*_I \otimes W_I$, where $\{W_I\}$ is an orthonormal basis of intertwiners for 
the four valent vertex on $\tau$ formed by the wedge boundaries (carrying their respective 
wedge spins $j_s$).\footnote{
If some of the incident wedges have spin zero then the 
valence of the vertex is effectively lower.}
The term in the cell amplitude (\ref{model:cellamp}) corresponding to
the spin assignment $\{j_s\}$ is thus a sum of spin network functions all having the same
spins on the edges of $[\di\n]^*$, given by the spins of the corresponding wedges, but different
intertwiners. If $\{\chi_\cg\}$ is an orthonormal basis of spin network functions then the
whole cell amplitude can be written as
\bearr
a_\n(\bfg_{\di\n}) & = & \sum_{\cg} \chi(\bfg_{\di\n})\:a_\n(\cg) \\
a_\n(\cg) & = & [\hat{C} \chi^*_\cg ({\bf h})]_{{\bf h} = \One}.           \label{netamp}
\eearr
$\bf h$ in (\ref{netamp}) is a connection on $\di\n$. The right side is evaluated on the trivial
connection in which every propagator is $\One$.

A spin worldsheet formulation of the model can now be defined using (\ref{netamp}) as explained
in \S \ref{sws1}. Actually (\ref{model_def}) gives the model in a form that is just a step
away from a spin worldsheet sum, via the methods of \cite{Rei94}. Each history assigns
spins to all the wedges of $\Dg^+$. This distribution of spins already defines a "spin surface",
consisting of the wedges carrying non-zero spin, each one coloured by its spin. 
These spin surfaces have open boundaries inside $\Dg$, they are generally just collections of
patches. However, when the connection is integrated out only spin distributions corresponding to
spin worldsheets in the sense of \S \ref{sws1} survive: The contribution of a spin distribution
to the ampitude of a spin network $\Cg$ on $\di\Dg$ is 
\be               \label{spindist_contrib}
\prod_{b\ \mbox{\scriptsize connection carrying edge of}\ \Dg^+}\int dg_b\ 
\chi_\Gamma^*\: w.   
\ee
with $\chi_\Cg$, the spin network function corresponding to $\Cg$, and $w$ the amplitude of
the history consisting of the spin distribution and the connection as given by (\ref{model_def}).
see (\ref{A_path_int}). The only spin distributions for which (\ref{spindist_contrib}) is non-zero
correspond to spin worldsheets spanning $\Cg$. For such a distribution (\ref{spindist_contrib})
is a sum over spin worldsheets all having the same, given, spins on each wedge,
but different intertwiners at the branch lines.



$sgn(s,\bar{s})$ is non-zero iff the pair of wedges $(s,\bar{s})$ intersects transversely in the
sense that they have no common tangent vectors (see (\ref{wedge_wedge_wedge}).  
From the expression (\ref{netamp}) one sees at once that if the occupied wedges 
(wedges with $j_s\neq 0$) in a 4-simplex contain no transversely intersecting pair then
the reduced amplitudes $\bar{\ag}_\n(\cg)$ are just those of BF theory. That is to say, the reduced
amplitude for a cell spin worldsheet consisting of a disk made of wedges none of which
intersect transversely is just $1$. The reduced amplitude for a branch line through the 4-simplex
would be $\dg_{I\,I'}$ where $I$ and $I'$ are the intertwiner labels at the point of entry and exit
of the branch line respectively. If we think of BF theory as a "free theory" from which GR is
obtained by adding an interaction, then in the spin worldshheet formulation the interaction occurs
at the transverse intersections of wedges. 

However, in a 4-simplex there are no cell spin worldsheets with branch lines
that don't also have transversely intersecting wedges. Similarly, all spin worldsheet vertices have
transversely intersecting vertices.
The only disks without transversely intersecting wedges are cones made of three wedges. 
Put another way, in a simplicial complex spin worldsheets have interactions in practically every cell
they enter. The only spin worldsheets without interactions are the minimal 2-spheres surrounding
1-simplices, which are the only 2-surfaces that can be made out of cell spin worldsheets that are
three wedge disks.

%
%
%

\end{document}